# Latent Chemical Space Searching for Plug-in Multi-objective Molecule Generation

Ningfeng Liu[1,2], Jie Yu[1], Siyu Xiu[1], Xinfang Zhao[1], Siyu Lin[1], Bo Qiang[1], Ruqiu Zheng[1], Hongwei Jin[1], Liangren Zhang[1], Zhenming Liu[1,3,*]

[1] State Key Laboratory of Natural and Biomimetic Drugs, School of Pharmaceutical Sciences, Peking University, Beijing 100191, China

[2] Peking-Tsinghua Center for Life Science (CLS), Peking University, Beijing 100871, China

[3] State Key Laboratory of Pharmaceutical Biotechnology, Nanjing University, Nanjing 210023, Jiangsu, People's Republic of China

[*] Corresponding Authors

## Abstract

Molecular generation, an essential method for identifying new drug structures, has been supported by advancements in machine learning and computational technology. However, challenges remain in multi-objective generation, model adaptability, and practical application in drug discovery. In this study, we developed a versatile 'plug-in' molecular generation model that incorporates multiple objectives related to target affinity, drug-likeness, and synthesizability, facilitating its application in various drug development contexts. We improved the Particle Swarm Optimization (PSO) in the context of drug discoveries, and identified PSO-ENP as the optimal variant for multi-objective molecular generation and optimization through comparative experiments. The model also incorporates a novel target-ligand affinity predictor, enhancing the model's utility by supporting three-dimensional information and improving synthetic feasibility. Case studies focused on generating and optimizing drug-like big marine natural products were performed, underscoring PSO-ENP's effectiveness and demonstrating its considerable potential for practical drug discovery applications.

# Main

Drugs have been discovered and deposited at a high speed in the last two decades. Until now, about 3,000 small molecule drugs have been approved according to DrugBank[1] and ChEMBL[2]. However, they are just a very limited amount in the whole chemical universe. The ChEMBL database currently lists more than $10^6$ compounds. Additionally, a virtual compound library named GDB-17[3], created by enumerating compounds within a restricted molecular size, records over $10^{11}$ compounds. It is estimated that the total number of potential drug-like compounds could ascend to an astonishing figure as high as $10^{60}$ units[4]. Meanwhile, Small molecules must meet a myriad of criteria to pass the clinical trials and gain commercial approval, such as drug efficiency, ADMET properties, structural novelty, and synthetic accessibility. These criteria significantly elevate the complexity, duration, and financial burden of drug discovery efforts[5], underscoring the critical need for more effective exploration of the chemical space.

Computer-aided drug design (CADD) assists chemists in many steps of the drug discovery pipeline such as target discovery, hit identification and, lead optimization[6,7]. It is poised to speed up chemical space exploration and address challenges in balancing drug efficacy with ADMET properties[8] or mitigating toxicity caused by off-target effects[7,9]. CADD can be divided into two main strategies: virtual screening and molecule generation (or *de novo* drug design). Although virtual screening has historically accelerated drug discovery, it still faces two critical challenges: (1) It can only apply to known compound libraries, meaning that it cannot independently and directly discover novel structures or scaffolds, leading to a bias when exploring the chemical space, and will need lots of effort on the structural optimization of hit to lead. (2) It usually can just be focused on one objective at once, and applying many layers of virtual screening to perform multi-objective screening will cause the efficiency of finding a perfect molecule sharply dropped. With the development of computers and artificial intelligence (AI), the efficiency of large variable space searching has been improved[10,11], providing significant potential for molecule generation in drug discovery. Compared with virtual screening, molecule generation has natural advantages in obtaining novel structures, parallelly optimizing multiple objectives, and relatively lower cost[12], thus making molecular generation more promising for exploring a large unknown chemical space[13].

While multi-objective molecule generation has become a hotspot in drug design, some challenges persist from previous studies[14-16]. Validity, FCD distance, logP, QED, and classical ligand-based QSAR models (such as targeting EGFR) are widely used as objectives, but many of them are away from practical demands for clinical use, making models validated on these toy objectives less reliable. This requires a model to have high flexibility in practical use, expecting a model to adapt to different multi-objective tasks with consistent reliability, which carries out differences in 'entrenched models' and 'plug-in models'[17]. Entrenched models tended to train molecule generation models with fixed datasets satisfying chosen properties. For instance, Kotsias used properties like activity, logP, and QED to form the input of an RNN model to generate molecules with these chosen properties[18], Gupta used datasets satisfying certain demands to fine-tune a pre-trained molecule generation model[19], Maziarka used paired molecules with both positive and negative properties to learn to generate molecules with same properties[20]. The disadvantage of these entrenched models was obvious that users could not directly put them into practical use when their tasks were not precisely met by the model, necessitating further cooperation and re-training. In contrast, the 'plug-in' model can perform generation with different combinations of objectives without additional training or adjustment. For instance, Yoshizawa used reinforcement learning with flexible cost function to guide the generation with a pre-trained molecule generation model[21], and Hoffman used zeroth-order optimization to guide molecular optimization in a pre-trained continuous latent space[22]. In 'plug-in' models, objectives (usually calculated directly or predicted by existing models) can be plugged in or out flexibly, facilitating practical application and enhancing result consistency since the reliability of each objective is dependent on the prediction model rather than the generation model itself. As an algorithm suitable for massive space searching, the particle swarm optimization (PSO) is a promising way to search the latent chemical space for small molecules[23]. Hartenfeller used PSO to perform molecule generation in discrete chemical space[24], but with the development of deep learning, it has become possible to apply PSO into continuous chemical latent space as demonstrated in Winter's work[25]. Specifically, PSO is used to guide several particles in a latent representation space formed by a pre-trained molecule-to-molecule model, and the decoder is used to get actual molecules in each PSO step for evaluation. In this paper, we further designed several variants of PSO to make it more suitable for molecule generation and drug discovery, and studied the performance of these variants. Meanwhile, we expand objectives to a comprehensive set of practical objectives by including a new easy-to-use target-based binding affinity prediction model, twenty ADMET prediction models. and several properties such as logP, SA scores, and similarity were included as objectives as well to achieve more functions such as molecular optimization. Evaluation results showed that our model can perform well in practical molecule generations and optimizations in up to 26-objective tasks.

## Results

### PSO variant design and performance study

In this paper, we introduced multiple PSO variants aiming problems in space validity, boundary, and distribution-searching, then evaluated them against a standard multi-objective task to identify a best variant for following experiments. We used the latent chemical space defined by a pre-trained SMILES-to-SMILES model CDDD[26], a 512-dimension space with a default domain of value (in the following paper it was abbreviated as 'default domain') between -1 to 1.

#### Variant design focused on initialization methods

In vanilla PSO, the particle position initialization was usually random, as in the default setting of PySwarms[27]. Though in its implementation in molecule generation, the latent chemical space constructed by the pre-trained model was not fully valid, which meant that not all points in the latent space could be decoded into valid molecules. By randomly sampling and decoding molecules 10,000 times from the default domain, we found that the valid rate in the default domain of the CDDD latent space was 30.28%, suggesting that employing the random initialization method (SpaceStart, S) resulted in over two-thirds of the particles failing to secure a valid initial position, thereby losing a large amount of information and sights of searching. To overcome this problem, we proposed a variant EasyStart (E), which sampled initial position by encoding random molecules from a molecule dataset, highly raising the chance of getting valid initialization points and expanding the initial sights without high time cost.

#### Variant design focused on boundary handle methods

In each step of PSO, particles moved to a direction by a certain velocity. When exceeding the boundary, a handle method was required to reposition the particle within the designated domain. In vanilla PSO (Nearest, N), particles returned to the nearest position in every exceeded dimension. But in a latent chemical space, particularly in CDDD, a particle beyond the default domain also had the chance to be decoded into a valid molecule, although the molecule would be repositioned back into a specific point inside the default domain after encoding. This suggested that points outside the boundary contained valuable information as well, and the Nearest methods might cause information loss. Therefore, we proposed ChemMapping (C), a method that firstly decoded the particle out of the boundary into a SMILES (Particles not valid were handled with the Nearest method), then encoded the SMILES back into the boundary, like mapping a point outside into a specific standard point inside. In this method, the particle retained structural information encoded by the latent space during the mapping process compared to the Nearest method.

**Variant design focused on the running process of PSO**

The third variant came from our study on the ability of the CDDD latent space to encode the structure-activity relationship (SAR) information. A basic assumption in SAR is that similar structures likely exhibit similar activity features, such as drug effects, binding affinity, even logP and metabolism activity. This assumption **suggested** that a latent chemical space encoding similar structures more closely than dissimilar ones (like clustering), would concentrate activity features used in molecule generation, meaning that a searching algorithm like PSO was likely to work better on these 'focused' latent chemical spaces. Thus, we tested this feature on the CDDD latent space. We randomly sampled 10,000 pairs of valid points, and their distribution is shown in **Figure 1A**. The Pearson correlation between the Euclidean distance and similarity was only -0.02276, suggested that due to the dimension disaster, most of the randomly selected distances were distributed between 17~20. In this range, the Euclidean-distance-based searching algorithms such as PSO were difficult to use SAR information effectively. Then we continued to randomly select pairs whose distances were between 0~17, and the distribution is shown in **Figure 1B** and **Table 1**, with a Pearson correlation valued -0.5345, showing a significant relationship between structure similarity and Euclidean distance. This suggested that in a small range, SAR could be effectively encoded in the latent space, and the smaller the range was, the higher the relationship between Euclidean distance and similarity. We also noticed another interesting result. In the study, 10000 pairs with distances ranging from 0 to 17 were randomly sampled. After discarding invalid pairs, 1052 valid pairs remained, with their median distance being 4.314 and 75% of them having distances less than 7.713. This meant that a valid point was more likely to gather more valid points in the CDDD latent space, and the 'decoding validity' was just like other molecular activity features that had a gathering potential in distributions. Moreover, the applicability of this characteristic to other latent chemical spaces deserves further investigation.

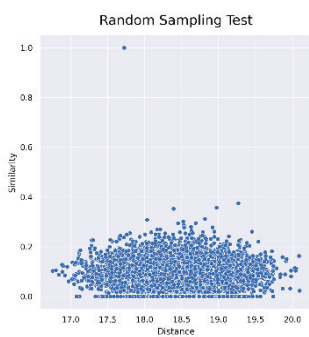

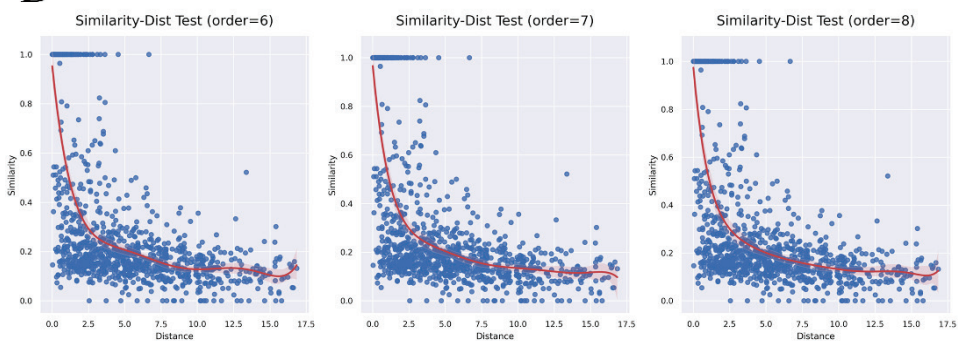

**Figure 1 Relationship between structure similarity and distance in CDDD latent chemical space. A: Randomly sampled in the whole space, B: Randomly sampled pairs with distances between 0 to 17**

**Table 1 Analytical results of point pairs randomly sampled from distance 0~17**

|  | Average | Min | 25% | 50% | 75% | Max |
|---|---|---|---|---|---|---|
| **Euclidean Distance** | 5.129 | 0.005718 | 1.777 | 4.314 | 7.713 | 16.85 |
| **Structure Similarity** | 0.3060 | 0 | 0.1304 | 0.1818 | 0.3111 | 1 |

The above results on the Euclidean distance ranging from 0~17 and no range limitation together formed a picture of the CDDD latent space, that activity features gathered like mountains and hills, in which were similar structures, while most of the latent space was plains, in which no significant relationship between structure and position shown. Thus for PSO, after the final step when every particle jumped according to a certain velocity, the swarm was likely to easily jump over the 'mountaintop', because the mountain (gathering area) was much smaller compared to plains (areas with no relationship between structures and positions). To overcome this problem, we proposed the third variant, wherein particles would spread out for a short distance after certain steps, and start the second round of the PSO process (**Figure 2**). We expected this method could help jump out and find the 'mountaintop' around the original convergence area. Therefore, we made particles randomly move +0.2 or -0.2 in each dimension centering the global best (gBest) position (Euclidean distance is 4.525) for spreading to make the process more even and suitable for finding the 'mountaintop'. In this paper, we designated the vanilla PSO running process as GlobalSearch (G) and our proposed two-stage PSO as PreciselySearch (P).

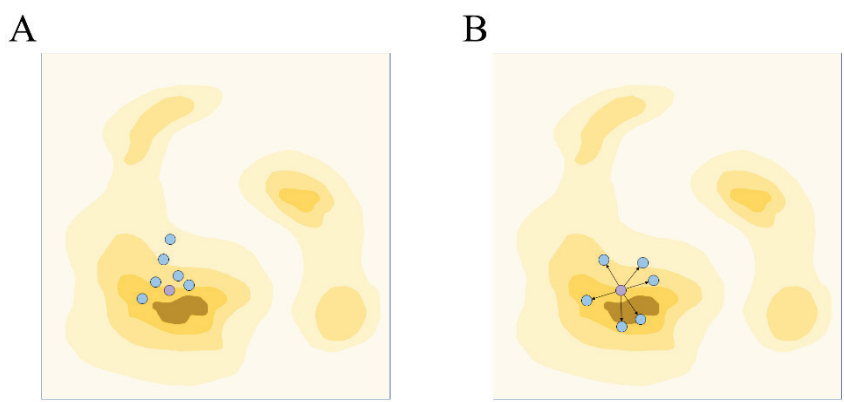

**Figure 2 The spreading operation. A: The end of the first stage, B: Particles spread out centering the gBest position. Blue points stand for particles, purple points stand for the gBest position.**

**Variant performance study**

In this paper, we performed a standard multi-objective molecule generation task to decide the best variant and tried to discuss the effect of each variant design. Aiming practical drug discovery, the standard task needed to include several essential and practical binding, drug-like, synthetic-related objectives, and should not include too many objectives making the importance of each perspectives of a drug hard to balance. Finally, for better discrimination, objectives having continuous outputs were better than those having classification outputs. Following these ideas, we defined the task as generating a molecule fitting four objectives: high binding affinity towards the PWWP1 domain of NSD3 (PDB: 6G2O[28]) predicted by a constructed model (see **Methods**), high drug-likeness (QED), MW between 200~600, and good synthetic accessibility (SA score), and the cost function was the sum of four objective scores with same weights (lower the better).

The overall comparison results are shown in **Figure 3A**, and grouped results are shown in following subgraphs. For initialization variants (**Figure 3B**), the result shows that EasyStart proposed in this paper was better than the traditional SpaceStart. This took advantage of more valid initial points covering more areas for future discovery. Among repeated experiments, the best cost scores (the value evaluating all objectives, see **Methods**) among particles after initialization in EasyStart ranged from -1.043~0.838 (median: -0.942), while in SpaceStart it ranged from -0.887~-0.487 (median: -0.743), proving that a better initial sight would be gained in the EasyStart method. For boundary handle variants (**Figure 3C**), the result surprisingly shows that Nearest was better than ChemMapping, meaning that the chemical-equivalent mapping operation could not gain better performance. One possible explanation was that although ChemMapping retained the information of the point/molecule itself, it lost too much positional context compared to the Nearest method. ChemMapping may perform a long-range jumping and set a zero velocity, potentially losing almost all information of structural similarity, and area/clustering information near the previous outside boundary position, which may do essential harm to searching algorithms with population decision functions as PSO. On the other hand, Nearest could keep this information, as well as keeping the robustness of the PSO running process. Although ChemMapping was not as promising as Nearest, **Figure 3D** shows that it had better performance than randomly resetting position when a particle was out of the boundary, indicating that the chemical information of the point kept in ChemMapping did have some utilizing potential in further studies. For PSO running process variants (**Figure 3E**), in 3/4 of all groups PreciselySearch showed significantly better performance than GlobalSearch, and in only one group it showed a slightly lower performance (notice that in PreciselySearch the step/epoch number was set to 20 (stage1) + 1 (spreading) + 20 (stage 2) = 41, and in GlobalSearch the step/epoch number was set to 41). This advantage came from the 'mountaintop' searching after the spreading operation, indicating that PSO tended to have lower improvement (even stacked in a fixed position quickly) during searching, and a spreading operation would force particles to re-search those areas they had jumped over. **Figure 3F** shows that stage 2 did improve the cost score to some degree. Finally, from **Figure 3A** we can conclude that the EasyStart-Nearest-PreciselySearch variant achieved the best performance among all model variants, and in the following paper we will name it as PSO-ENP and use it for further studies.

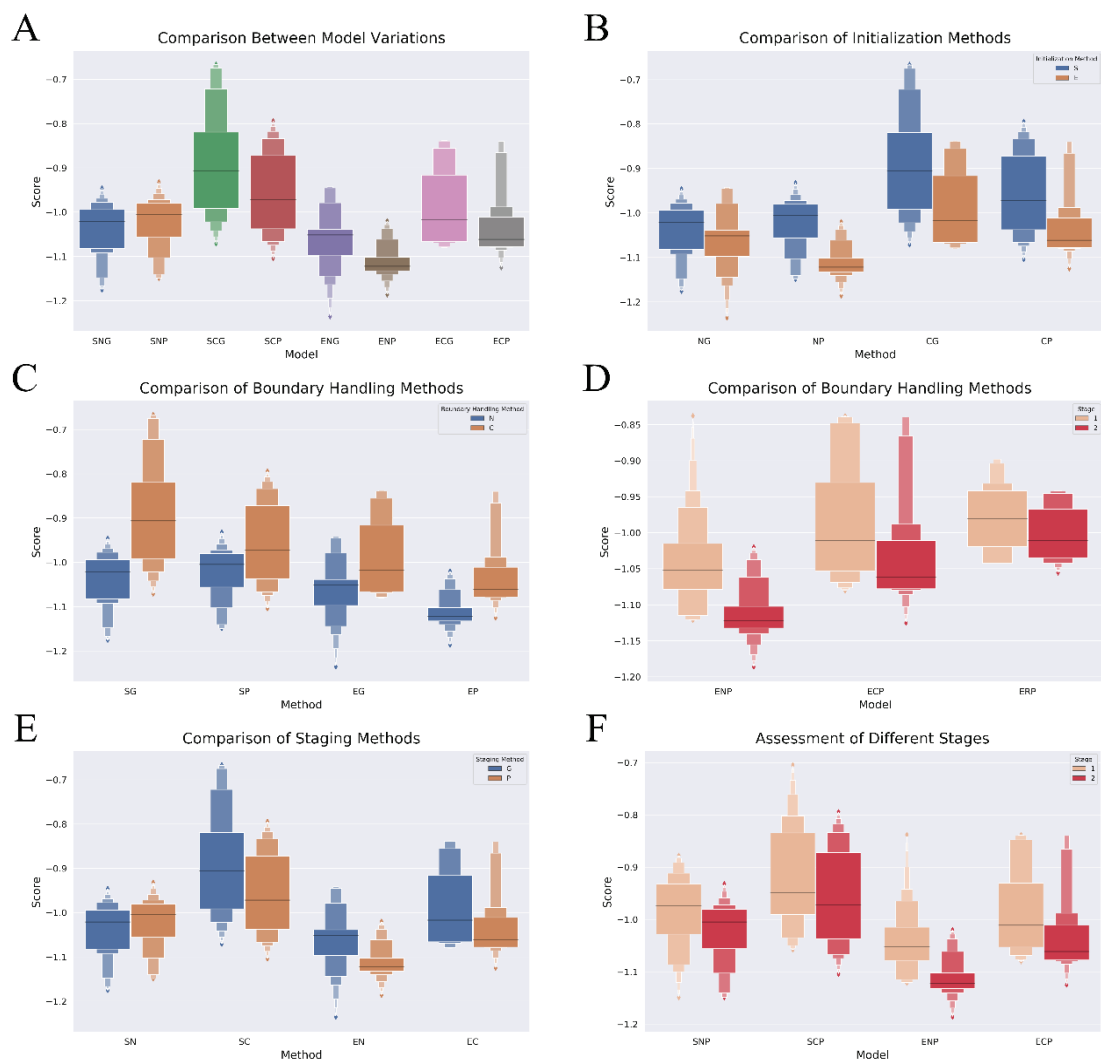

Figure 3 PSO variant comparison in a standard multi-objective task. A: The overall results. B: Comparison grouped by initialization methods. C: Comparison grouped by boundary handling methods comparison. D: Boundary handling methods comparison in fixed variant settings. E: Comparison grouped by staging methods. F: Comparison between stage 1 and 2

**Single-objective molecule generation study**

Single-objective aimed to generate molecules satisfying only one objective. They were not suitable tasks to indicate a model's actual value in drug discovery, but could reflect a model's ability to deeply search the latent chemical space for 'extreme molecules'. **Table 2** shows the result of PSO-ENP in six single-objective tasks (SA aims for lower values, others aim for higher values), suggesting that PSO-ENP could successfully find extreme molecules encoded in the latent space with a promising searching ability.

**Table 2 Extreme single-objective molecule generation results**

| Objective | Case generated molecule | Average value |
|---|---|---|
| logP | 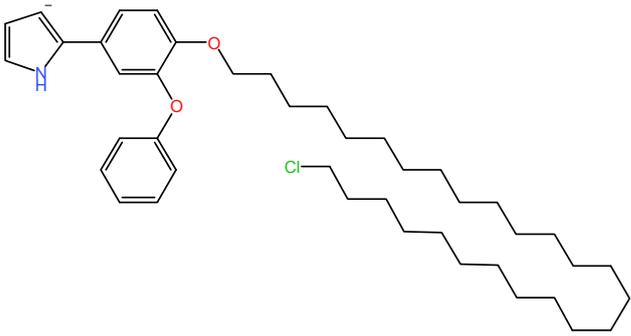 | 11.094 |
| HBA | 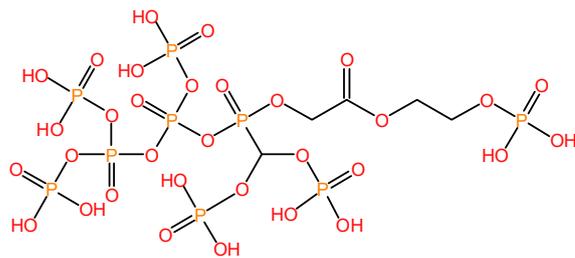 | 23.100 |
| HBD | 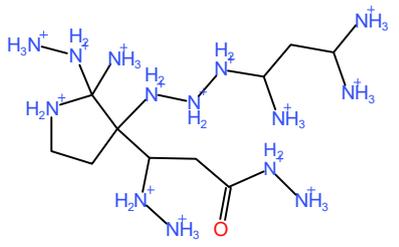 | 21.200 |
| Fluorine atom numbers | 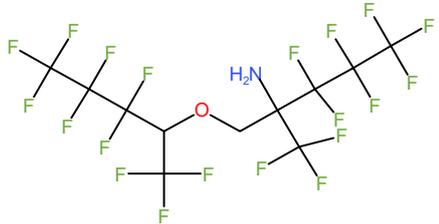 | 8.300 |
| Ring numbers | 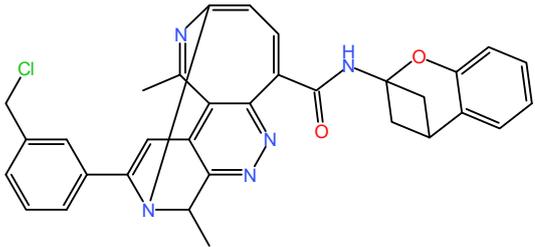 | 7.750 |
| SA | 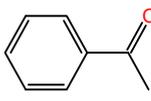 | 1.332 |

One essential objective in drug discovery is binding affinity towards a certain target. The best/average cost score of the population's curves of PSO-ENP in NSD3 or EGFR (PDB 2ITY[29]) binding single-objective tasks are shown in **Figure 4**, and the final generated molecules have predicted binding affinities ($K_d$) of -1.103±0.762 (NSD3) and -0.930±0.641 (EGFR). From the best score curves, we can see that the model could continuously search for better positions, and from EGFR we can observe a significant improvement after the spreading operation (epoch 21), and the effect does not stop after the operation, but keeps improving the score in a long period. This indicates that the spreading operation benefited not only from the new position, but also from a better and wider searching area for the following searching. From the average score curves, we can observe the same trends in stage 1 (before epoch 21), but a significant drop can be observed after the spreading operation. This was not an unexpected result, because there should be only one 'mountaintop' area around the original position before spreading, making most of the particles spread into a worse position, which damaged the average score. As a result, this observation proves the functionality of the spreading operation to find the 'mountaintop'. A docking and binding pattern prediction case is shown in **Figure S1**, proving that PSO-ENP could generate molecules having promising binding potential towards given targets.

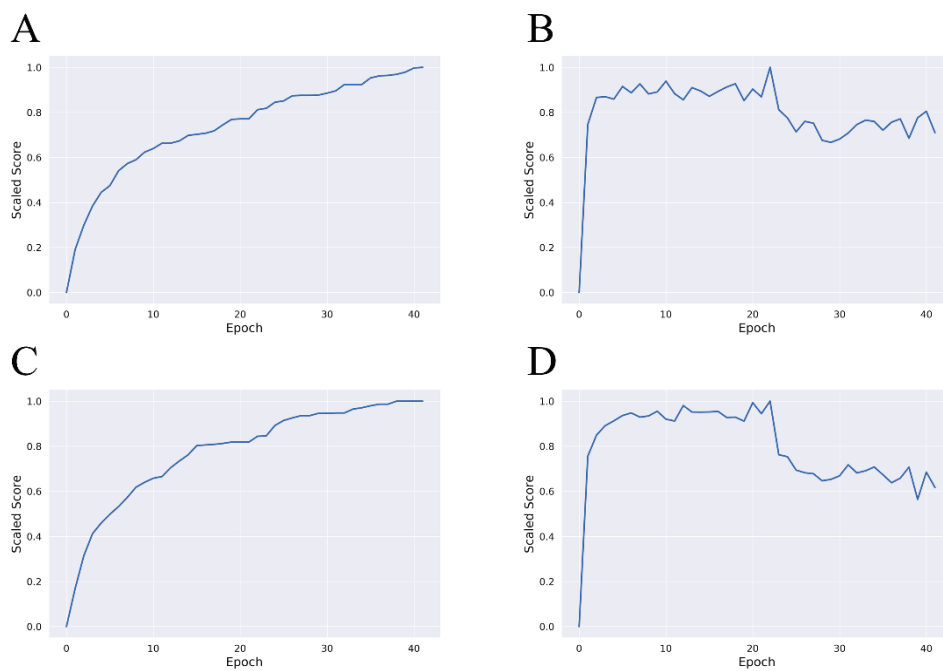

**Figure 4 The binding affinity objective values during the iteration. A/C: gBest values (repeated experiments' average) for NSD3/EGFR. B/D: Particles' average values (repeated experiments' average) for NSD3/EGFR**

Finally, a baseline test GuacaMol[30] was done, and the results of single-objective tasks are shown in **Table S1**. Although promising performances were observed in the result, it is important to note that many tasks in the GuacaMol test were not frequently used in drug discovery. In practical use, objectives related to binding affinity, drug-likeness, ADMET and synthetic accessibility were more concerned, though GuacaMol lacked target-based binding affinity tasks, and included too many single or multi-objective tasks including similarity-related objectives. As a result, the GuacaMol test in this paper served merely as a reference and was not considered as an essential assessment.

## Multi-objective molecule generation study

### Performance study for equal-weighted multi-objective tasks

The final goal of our model was to perform multi-objective molecule generation aiming at practical tasks, thus we studied the model's performances in different settings of tasks. We designed (See details in Methods) a 3-objective task (binding, QED, SA), a 6-objective task (binding, logP, HBA, HBD, MW, SA. The idea was to split QED into different perspectives), and a massive 26-objective task (6-objective task + QED + 19 ADMET objectives. BBB was removed from ADMET objectives for it was a high variable objective for different demands and was related to many other objectives such as logP and QED). We first conducted experiments with all objective weights set to 1, and the results are shown in **Figure 5** (binding to NSD3) and **Figure S2** (binding to EGFR).

From the best score curves (**Figure 5ACE**), it is observed that the cost score (red) is improving continuously during PSO-ENP running processes, and objective scores can also keep improving overall. Notice that the SA score is an exception, for binding affinity requires structural features and some complexity, which conflict with the synthetic accessibility. Similarly, the average score curves (**Figure 5BDF**) demonstrate the same trend of an overall improvement in Stage 1, and comparing the end of Stage 2 to the initial score, an improvement is also observed. These indicate that the PSO-ENP model could optimize multiple objectives parallelly during the searching process, finding better molecules in a large latent space. Besides the overall conclusion, other interesting findings can be seen in **Figure 5**: (1) The binding affinity was a complex objective compared to others, though it can be observed from the best score curve that the binding affinity curve seems more similar to the total cost curve than other objectives (especially in the 3/6-objective task). This may come from that binding to a pocket requires specific structural features, which might be encoded better in a chemical latent space pretrained by structural representation such as SMILES or molecule graphs. Compared with SA or QED, binding affinity may have a more significant hotspot in the space leading to an easier search, especially in a space having SAR encoded well (as discussed in Variant design focused on the running process of PSO). (2) The effect of the spreading operation can be significantly observed in both best and average score curves. As discussed before, this result shows that the spreading operation also demonstrated efficacy in multi-objective tasks. (3) With the growth in the number of tasks, the score curves seem more unstable and divergent. It is reasonable to assume that an increase in the number of objectives might dilute the models attention on individual objectives, causing a more random optimization in each objective. Notice that a constant count of 20 particles was employed for all multi-objective tasks, meaning that on average a particle needed to pay attention to more than one objective in the 26-objective task. Another thing we noticed is that with the growth in the number of tasks, the performance of binding affinity optimization dropped significantly (especially can be observed in average score curves), also causing the loss of its similarity to the total cost curve in the 26-objective task. This was simply because although the binding affinity score contained more robust structural features, more tasks still lowered its influence on the total cost score. From the perspective of final binding affinity prediction results, a drop could also be seen during the task number's growth, though in all tasks PSO-ENP could successfully generate promising molecules binding the target protein. **Figure S3** shows the predicted binding patterns of case molecules generated in 3/6/26-objective tasks. Agreeing with Böttcher's study of NSD3[28], case molecules generated by PSO-ENP could utilize key residue SER314 to form H-bond, and utilize key residue TYR281 and PHE312 to form π-π interactions, showing a practical binding potential.

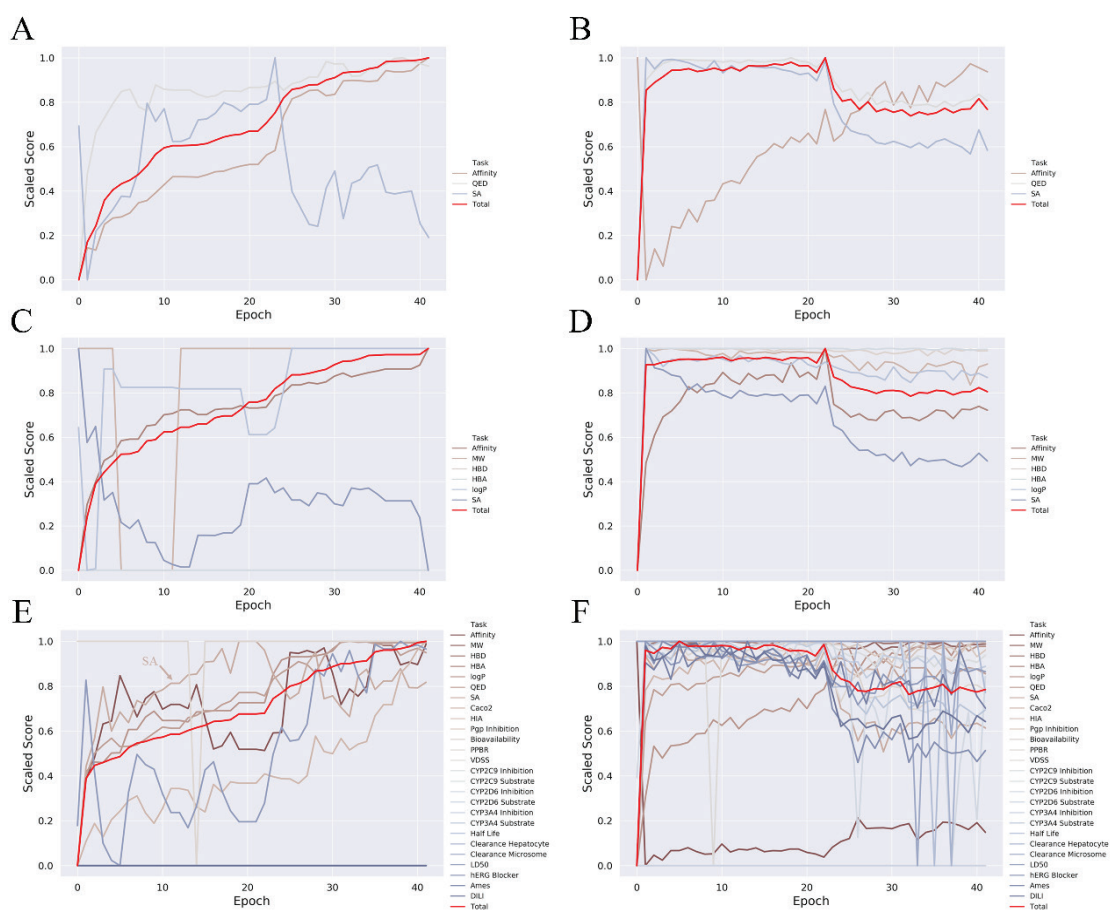

**Figure 5 The multi-objective task scores during the iteration. A/C/E: gBest values (repeated experiments' average) for 3/6/26-objective tasks. B/D/F: Particles' average values (repeated experiments' average) for 3/6/26-objective tasks**

The above discussion which focused on optimization performance during epochs can reflect the model's ability in space searching, though in practice we are more concerned about the final changes and results. For example, while an objective might achieve its best score after initialization, and then slightly drop in the following search process, maintaining a high level throughout signifies successful generation as well. Thus, we used Optimization Rate (OR, the proportion of objectives been successfully optimized) and Success Rate (SR, the proportion of objectives been successfully satisfied in the output molecule) to evaluate the ability of practical molecule generation (details can be found in Methods), and the results are shown in **Table 3**. In the 26-objective task the generated (best) molecules optimized over 75% of objectives, over 90% of generated molecules had at least half of objectives optimized, and the population particles optimized over 90% of objectives. Among all tasks, the particle population in all repeated experiments successfully optimized at least half of the objectives. From the SR value, we can see that almost all objectives could be satisfied in the generated molecules in all tasks. Notice that values like SR show a trend that with the growth of the number of objectives, the value grows higher as well. This was probably because ADMET values contained many 0/1 classification objectives which were easy to gain and keep. The above results overall indicate that PSO-ENP had the ability to generate molecules satisfying most of the objectives required, and had a promising potential to be put into practical use. At last, suitable GuacaMol's multi-objective tasks were also performed and the results are shown in **Table S2** and **Figure S4**.

Table 3 Optimization rates and successful rates of multi-objective tasks

| Objective | Target | $OR_{best}$ Average% | $OR_{best}$ Half Rate% | $OR_{pop}$ Average% | $OR_{pop}$ Half Rate% | SR Average% |
|---|---|---|---|---|---|---|
| 3 | NSD3 | 71.7 | 80.0 | 75.0 | 100.0 | 81.7 |
| | EGFR | 71.7 | 85.0 | 81.7 | 100.0 | 86.7 |
| 6 | NSD3 | 55.8 | 90.0 | 100.0 | 100.0 | 96.7 |
| | EGFR | 61.7 | 95.0 | 100.0 | 100.0 | 98.3 |
| 26 | NSD3 | 76.2 | 90.0 | 91.7 | 100.0 | 94.8 |
| | EGFR | 80.0 | 100.0 | 93.6 | 100.0 | 96.2 |

**Performance influence study in different weight settings**

As discussed above, more tasks would have negative effects on binding affinity optimization. Considering that the binding affinity objective was usually the most concerned in drug discovery, we discussed the weight-changing effect in the following paragraph. We followed the same experiment settings as the previous 26-objective task, but varied the weight of binding affinity objectives in 1.0/2.0/4.0/8.0. The average score curves are shown in **Figure S5**, and the OR and SR results are shown in **Table S3**. The success rates of the generated molecules in four weight settings for binding affinity objectives were 0%/0%/15%/45% (NSD3) and 5%/15%/45%/90% (EGFR), suggesting that a higher weight could improve the ability of the binding affinity optimization, and this can also be concluded from average score curves. However, overall the optimization of all tasks seemed to be more unstable with higher binding affinity weight, and the OR or SR dropped as well. This came from the same reason of more objectives causing worse overall optimization process for other tasks' attention has been taken to the over-weighted objective. Finally, we should notice that despite the performance of other tasks declined in certain areas, SR values were still kept at a high level (over 90%), indicating that setting higher weight to binding affinity (or other more concerned objectives) is generally viable in practical drug discovery.

**Performance study of molecule optimization**

As a side function of multi-objective molecule generation, the PSO-ENP could also be used to perform molecule optimization, only needing to add an objective restricting the similarity between searched molecules and the reference molecules. For this study, we temporally added a variant called 'PSO-ENP-0' which initialed one particle as the reference molecule, to include the reference molecule's structural information from the beginning, instead of masking this information and guiding particles with only similarity scores. We used GuacaMol baseline tasks for this study, as they aimed to improve molecule properties with similarity restrictions, and the results are shown in **Table 4**, and the case molecules are presented in **Figure S6**. We can see that PSO-ENP(-0) had a promising ability to optimize structures, and in most of the tasks, not including the reference structural information seemed to perform better. This indicates that the PSO-ENP could stand-alone and use its searching ability to find the structural similarity area without telling it the exact area at the beginning. This also resulted in large scaffold-hopping compared to PSO-ENP-0 when optimizing Amlodipine (**Figure S6**), for it may discover various areas with similar structures, but may not discover the area with a similar backbone or scaffold as well. This suggests that when doing molecule optimization using PSO-ENP in practice, it is needed to decide whether to include the exact structural information depending on whether the task needs scaffold-hopping or just sidechain optimization.

Table 4 Baseline multi-objective optimization tasks. Bold numbers are the best values of tasks

| Task | Discription | Graph MCT | PSO-ENP | PSO-ENP-0 |
|---|---|---|---|---|
| Osimertinib MPO | Optimize TPSA and logP of Osimertinib with similarity restrictions | 0.784 | **0.825** | 0.529 |
| Fexofenadine MPO | Optimize TPSA and logP of Fexofenadine | 0.695 | **0.720** | 0.612 |
| Ranolazine MPO | Optimize TPSA, logP and the number of F atoms of Ranolazine | **0.616** | 0.600 | 0.573 |
| Perindopril MPO | Optimize aromatic ring number of Perindopril | 0.385 | **0.511** | 0.472 |
| Amlodipine MPO | Optimize ring number of Amlodipine | 0.533 | 0.565 | **0.755** |

**Drug-like marine natural product analogue generation and structure optimization**

Studies of marine natural products (MNPs) have accelerated with advancements in marine and extraction technologies, resulting in the rapid deposition of new MNPs at a rate exceeding 1,000 per year, indicating that MNPs are promising sources of drug discovery. Previous works[31] have conducted cheminformatics analyses based on the CMNPD[32] database, identifying drug-like large MNPs and suggesting that MNPs generally exhibit lower drug-likeness and synthetic accessibility due to their complex and sizable structures. Furthermore, brominated MNPs are posited to hold significant value for drug discovery. Building upon this foundation, this study undertook two case studies: (1) Generation of drug-like large MNPs analogues; (2) Optimization of an MNP for enhanced target binding affinity, drug-likeness (QED), Synthetic Accessibility (SA), whilst retaining its bromine structural feature.

**Case study 1: Drug-like big MNPs analogues generation**

Employing the Tanimoto similarity function and large MNPs catalogued in both DrugBank and CMNPD as described in the previous work[31] as the objective and reference molecules, we aimed to generate drug-like large MNPs analogues. The resultant molecules, depicted in **Figure S7**, maintained the structural features of MNPs, characterized by long chains or large rings, and exhibited considerable structural diversity. Additionally, we assessed seven key drug-like attributes of entities from DrugBank, CMNPD, large MNPs in DrugBank (MNPxDrug), and the newly generated analogues. PCA was conducted with six drug-like properties (excluding QED), and the results are shown in **Figure S8** and **Figure S9**. These results demonstrated that the generated molecules possessed more concentrated property distributions, highlighting the PSO-ENP models capability to distil and intensify Structure-Activity Relationship (SAR) information from a singular similarity objective within the CDDD latent chemical space. The generated molecules raised the QED of MNPxDrug while keeping its features of structural complexity, which can also illustrate the same conclusion. The enhancement in QED among MNPxDrug molecules, whilst preserving structural complexity, corroborated this conclusion. As in practical use, this indicates the PSO-ENP models utility in data augmentation and ligand -based molecule generation for identifying promising bioactive molecules.

**Case study 2: MNPs structure optimization**

The methotrexate binding site of Dihydropteroate Synthase (DHPS) is widely studied to avoid drug resistance. We firstly performed a virtual screening to dock CMNPD molecules to DHPS (using PDB: 2H23[33], with predicted binding patterns illustrated in **Figure 6A**), and selected the highest ranked compound as the reference molecule (**Figure 6B**) for optimization. Optimization efforts prioritized similarity as a structural constraint (weighted at 8), aiming to elevate QED and SA while maintaining high binding affinity and at least one bromine atom, and the generated molecules are shown in **Figure 6CDEF**. According to Hevener[34] and Zhao's[35] works, ligand binding at this site predominantly involves hydrogen bonds with ASP101, ASN120, ASP184, LYS220, a hydrogen bond or salt bridge with ARG254 (also serving as a stacking platform), and Van der Waals force with PHE189. From the result we can observed these patterns, and the original ligand of 2H23 and the reference MNP each occupying a different cavity. Surprisingly, molecules optimized by PSO-ENP demonstrated the capability to engage both cavities, thereby establishing more selective, specific, and stable binding configurations. Additionally, these optimized entities exhibited a diverse range of interactions with ASP101, PHE189, LYN220, and ARG254, aligning with findings from prior studies. Owing to a suboptimal angle of indole ring systems, Molecule E did not fully occupy the cavity on the right, though forming a hydrogen bond with LEU26 and GLY63 as compensation to stabilize its binding in a half-open space.

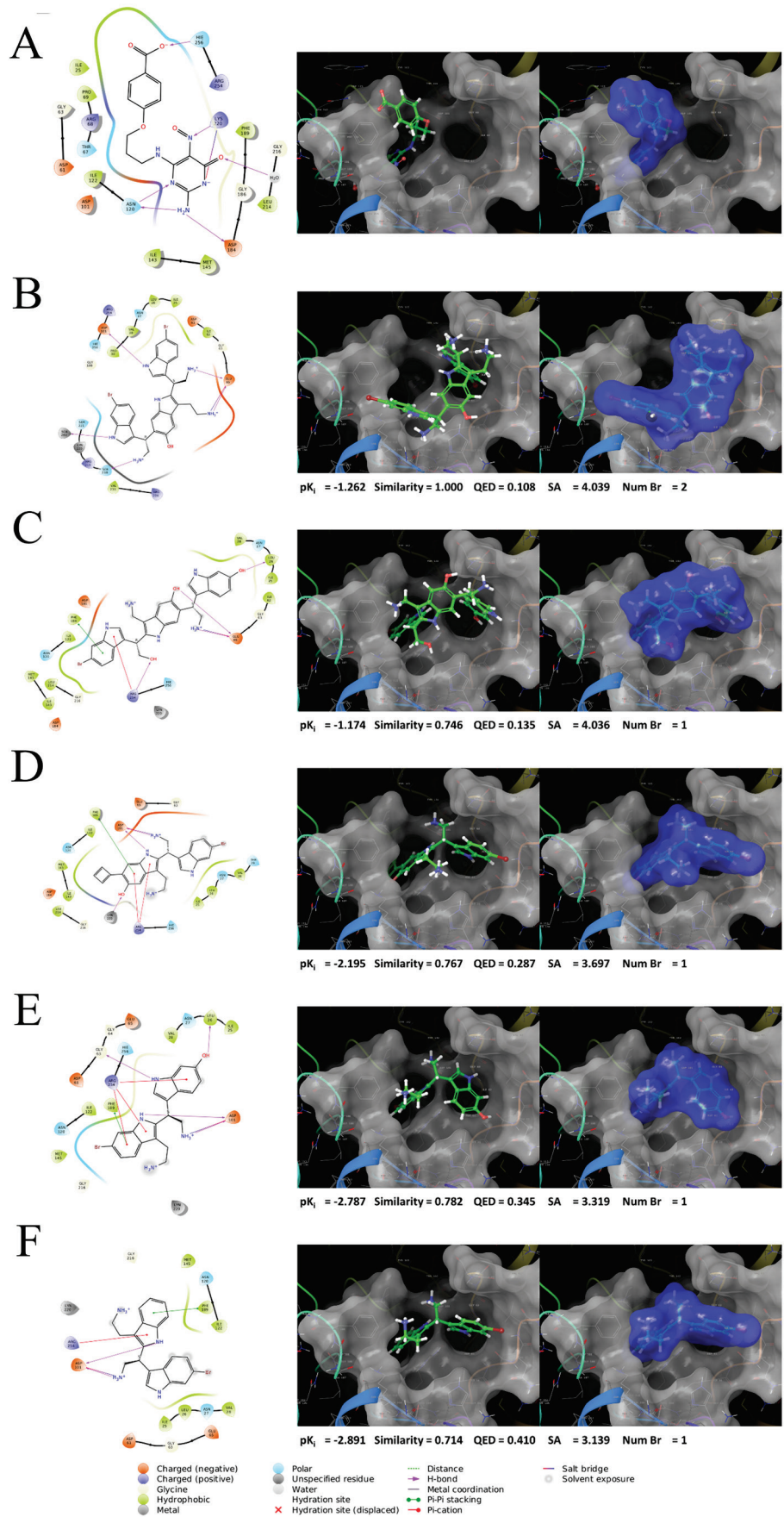

**Figure 6 The molecular binding patterns of optimized MNPs towards DHPS. A: The original**

**ligand of 3H23. B: The reference MNP given by virtual screening. C-F: Case molecules optimized from the reference MNP.**

From the perspective of optimization, four optimized molecules have high similarity to the reference molecule, including analogous backbones of indole ring systems. Although all predicted binding affinities were maintained at elevated levels, notably, only Molecule C, retaining all three indole ring systems, demonstrated a superior predicted binding affinity compared to the reference molecule. Molecules D, E, and F exhibited a reduction in one indole ring system, thereby decreasing structural complexity and significantly enhancing QED and SA, indicating a reasonable optimization thinking embedded within the search algorithm. At last, each of the optimized molecules successfully retained at least one bromine atom. These results showed the PSO-ENP models capacity for effective multi-objective optimization, even when addressing molecules of considerable complexity.

## Binding affinity and ADMET prediction models

### Performance study of the binding affinity prediction model

The binding affinity prediction model used in this paper was constructed with a two-way 3D-pharmacophore-based deep neural network described in Methods. The adoption of the SMILES representation alongside a novel model constructed specifically for this study aimed to ensure direct integration of the CDDD representation into this model, thereby minimizing potential confounders in the performance evaluation. An ablation study, detailed in **Figure S10** (refer to Methods for more information), confirmed that each module of the final model (termed Pharm3D-DTA) is essential. The results of 5-fold cross-validation, along with the test set outcomes, are presented in **Table 5** and **Figure 7**, demonstrating fine and robust prediction performance.

**Table 5 The 5-fold validation result of the final model Pharm3D-DTA. '_r' represents the value range across 5 folds.**

|  | RMSE | RMSE_r | R2 | R2_r | Pearson | Pearson_r |
|---|---|---|---|---|---|---|
| **Train** | 0.7543 | 0.7194~0.7957 | 0.8513 | 0.8353~0.864 | 0.9245 | 0.9169~0.9307 |
| **Valid** | 1.3332 | 1.3007~1.3551 | 0.5349 | 0.5185~0.5537 | 0.7367 | 0.7224~0.7494 |

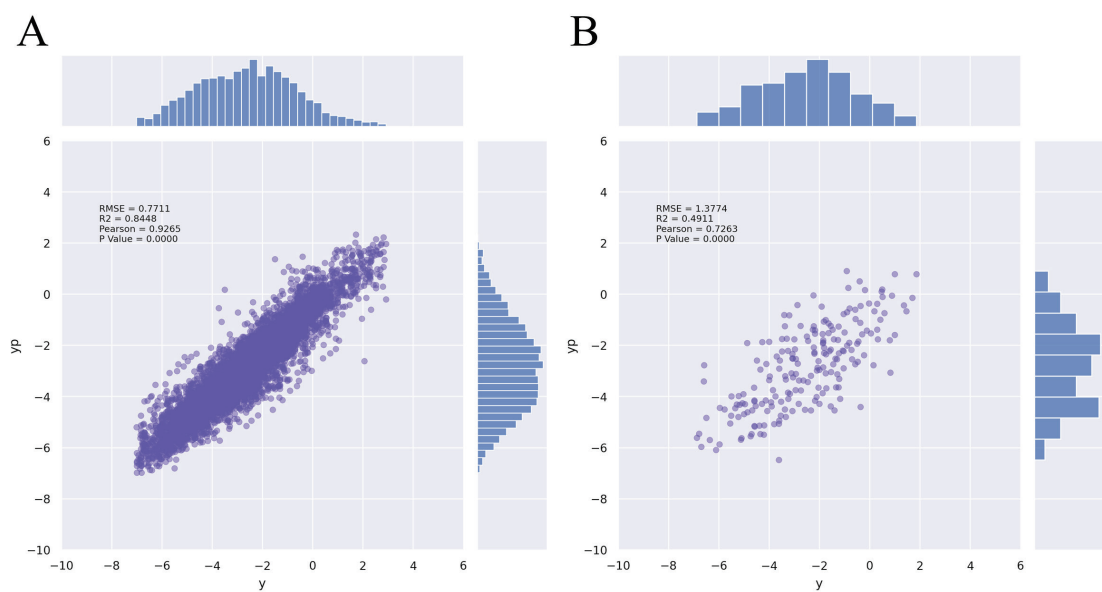

**Figure 7** The binding affinity prediction performance of Pharm3D-DTA. A: Train. B: Test

**Performance study of ADMET prediction models**

The ADMET prediction model presented in this study was developed using various traditional machine learning techniques, through an extensive search for optimal performance across different methods and hyperparameter settings. The final results for all thirteen classification prediction models and seven regression prediction models are detailed in **Table S4** and **S5**. The majority of ADMET models demonstrated considerable predictive capabilities. Models targeting features such as CYP2C9, vdss_lombardo, and half_life_obach exhibited relatively lower performance metrics, yet with AUC values above 0.5 and $R^2$ values exceeding 0, indicating their efficacy in discriminating between different molecules. Consequently, these models were retained for molecule generation studies.

## Conclusion

In this study, we introduced three modifications to adapt PSO for the exploration of latent chemical space and compared the performance of these variants in a standard and practical task. The findings indicate that each modification confers certain advantages to the PSO process, and the method of initializing from existing molecules or incorporating a spreading operation outperformed traditional PSO methods. Additionally, we developed a target-based binding affinity prediction model,

Pharm3D-DTA, and ADMET prediction models, alongside several structural and drug-likeness calculators, to offer versatile options for users in practical drug design. Finally, the superior model variant, PSO-ENP, was evaluated across various single-objective and multi-objective tasks encompassing up to 26 objectives, and was further applied in two case studies to generate drug-like large MNP analogues and optimize an MNP. These applications demonstrate the models promising capability to navigate the latent chemical space, optimize multiple objectives concurrently, generate molecules meeting the majority of objectives, and effectively handle the generation or optimization of complex MNP structures.

## Methods

### Datasets and data preprocessing

For the training of the binding affinity prediction model (Pharm3D-DTA), we use PDB-Bind 2020 Refined datasets[36], which includes 5316 protein-ligand complexes of high structure and assay qualities. For target proteins, PyMOL (https://pymol.org) 2.4.1 was used to process the raw files for target proteins, whereas RDKit was employed to convert the raw files of ligands into SMILES format, both discarding any invalid entities. Subsequently, complexes were correlated with assay binding affinity values (Ki or Kd, in nM), resulting in a dataset of 5,237 paired entries for the training of the binding affinity prediction model.

For the ADMET prediction models training, a ligand-based approach was adopted, aggregating 20 datasets representing various ADMET properties from the Therapeutics Data Commons[37] (TDC) and converting them into ECFP4 fingerprints via RDKit.

### Generation model's structure and variant design

Initially, a pretrained model, CDDD[26], served as the foundation of the latent chemical space, wherein each SMILES representation was encoded by CDDD into a 512-dimensional vector, representing a position within a 512-dimensional space. Each position within this space was decoded back into a unique SMILES-like string by CDDD, if valid, transformed into corresponding molecules. The study noted that the encoded representations were confined to a range of -1 to 1, termed as the 'default domain'. However, vectors extending beyond this domain could also be decoded by CDDD, potentially resulting in valid molecules. The molecule generation framework introduced in our study employed Particle Swarm Optimization (PSO) for navigating the latent chemical space, similar to the approach proposed by Winter et al.[25], albeit incorporating several variant designs to enhance the framework. The framework was articulated as follows:

(1) A cost function was established based on selected objectives, with the intent of generating molecules fulfilling these criteria. The cost score of a molecule was:

$$Cost\ Score = \sum_{i=1}^{n} w_i s_i$$

where n was the number of objectives, $s_i$ and $w_i$ were the score and weight of the $i^{th}$ objective of a molecule. This value is designed to be better when smaller.

(2) Positions of a population of particles was initiated in the latent space. These particles were used to move and search in the latent chemical space, decode positions to molecules, and calculate cost scores.

(3) Particles decoded their positions using CDDD to generate a molecule and then computed the cost score through direct calculation methods (e.g., MW calculated using RDKit) or prediction models (e.g., binding affinity or ADMET prediction models). If a position failed to be sampled as a valid molecule, the cost score was assigned a value of 10,000, with each objective score assigned a value of 100.

(4) PSO was applied to update optimization steps. Firstly, pBest recorded each particle's 'best position', achieving the best cost score in each particle's history steps. Then, gBest recorded all particles' 'best position', achieving the best cost score in all particle's history steps (global best position). Then for each particle at step t, a velocity of this step was calculated:

$$v_t = wv_{t-1} + c_1 r_1 (pBest - x) + c_2 r_2 (gBest - x)$$

where x represented the position of the particle, w, $c_1$, $c_2$ were weights for inertia, individual learning, and population learning, $v_{t-1}$ was the velocity of the last step, and $r_1$ and $r_2$ were weights randomly ranging from 0 to 1 in each calculation. Specifically, the initial velocities were all set to 0. After velocity calculation, each particle updated its position (moving) following:

$$x_{t+1} = x_t + v_t$$

where $x_{t+1}$ and $x_t$ were positions of the next step and this step of the particle.

(5) The iteration looped between (3) and (4) until reaching the max iteration number, then gBest's decoded molecules were regarded as the output of the molecule generation.

Several variants were designed in this study:

(1) EasyStart used the same preprocessing methods as was applied to PDB-Bind 2020 refined, to preprocess PDB-Bind 2020 general dataset's ligands. Then, EasyStart utilized this processed data as a source to sample SMILES strings, encoded via CDDD to initialize the positions of particles.

(2) ChemMapping was a chemical-equivalent boundary handling approach. Upon a particles excursion beyond the confined domain, it was initially decoded via CDDD, followed by a validation check of the decoded string. If the string was valid, it was re-encoded by CDDD back into the confined domain. If invalid, the particle was repositioned within the domain using the Nearest methods, adjusting each out-of-bounds dimension to its nearest boundary limit (either 1 or -1).

(3) The PreciselySearch variant executed a spreading operation subsequent to specified iteration steps, by randomly adjusting each dimension of each particle around the gBest position by either +0.2 or -0.2. This adjustment fixed the spreading distance at 4.525, ensuring a uniform distribution. After the operation, the velocities of all particles were reset to zero, pBest was updated to the current position (marking the initiation of the next round of PSO), and gBest was preserved. The default domain was set to a range of -1.4 to 1.4, preventing particles from exiting the boundary after spreading. A second round of PSO was then conducted, with the output of the molecule generation process being provided after a certain number of iterations.

Following evaluation, the PSO-ENP variant, integrating EasyStart, Nearest, and PreciselySearch, was demonstrated superior performance and was designated as the definitive model. The structure of this model is depicted in **Figure S11**.

**Experimental settings in the evaluation of molecule generation models**

For variant comparison tests, the experimental framework employed involved 20 particles, conducting 20 iterations for each of the two stages within the PSO-ENP protocol. The initialization of particles was conducted by setting the random seed within the range of 0 to 4, and for each initialization scenario, four repeated generations were performed, culminating in a total of 20 replicated experiments for each variant under consideration. In the case of the GuacaMol baseline assessment, the standard procedure also incorporated 20 particles and 20 iterations for each of the two stages within the PSO-ENP protocol, with the random seed set within the range of 0 to 99 to facilitate the generation process on 100 occasions. Some exceptions were rediscovery tasks, involving the use of 250 particles, 70/30 steps for each stage, and generating 10 molecules, to strike a balance between performance efficiency and time cost. The parameters for single-objective and multi-objective assessments remained consistent with those delineated for variant comparison tests. The optimization rate (OR) was delineated as follows:

$$OR = \frac{n_{positive}}{n} \times 100\%$$

where $n_{positve}$ represented the number of objectives having positive optimizations in comparison to their post-initialization values, and n denoted the number of objectives changed relative to their initial values. The success rate (SR) was defined as the proportion of objectives attaining certain goals, with the goal for each objective specified as follows: binding affinity (-log nM) > -4, MW < 500, HBD ≤ 5, HBA ≤ 10, 0 < logP < 5, QED ≥ 0.5, SA < 4, Caco2 (log cm/s) > -5.15, HIA > 30%, Pgp no inhibition activity, bioactivity ≥ 20%, PPBR < 90, 0.04 < VDSS < 20 (L/kg), CYP2C9/CYP2D6/CPY3A4 no inhibition or substrate activity, half life ≥ 3h, microsome/hepatocyte clearance ≥ 5 ml/min/kg, -log LD50 < 2.660 (predicted with the origin 6G2O ligand) or 2.961 (predicted with the origin 2ITY ligand), hERG no blocker activity, Ames negative, DILI negative.

**Calculator construction**

A variety of calculators, designed to provide objective scores requiring no additional training, were constructed directly. The QED, logP, TPSA, HBD, HBA, nRB (number of rotatable bonds), and MW (molecular weight) were computed directly utilizing RDKit. The SA score was calculated by sascorer[38], and similarity was calculated through Tanimoto similarity using 1024-bit ECFP4 given by RDKit.

**Binding affinity and ADMET prediction models' construction and evaluation**

Prior to the training of the binding affinity prediction model, the input derived from the targets pdb file was preprocessed by the following steps enlightened by Desaphy's work[39]:

(1) Using pocket file and ligand file from the PDB-Bind, calculating the ligand's geometry center and generate a cube grid centering it. The grid has a resolution of 1.5 Å with 14 grid cells in each length, total 2,744 cells in a grid.

(2) Then we assigned each cell to a pharmacophore feature. Cells within 2.5 Å from a protein atom is assigned to 'IN'. For rest of the cells, calculate:

$$t = \sum_{x<R} \frac{1}{x}$$

where R was set to 4 Å, x represented the distance from a cell to the nearest 'IN' cells, and if t<1/T (T was set to 1.75 Å), then the cell was designated as 'OUT' (meaning that they were too far away from pocket areas). Cells with fewer than three unassigned neighboring cells were also marked 'OUT' (for it may be too alone for useful features). Cells within a distance greater than R to the closest protein atom were categorized as DU. (not possibly being useful). Remaining cells were classified according to the closest protein atom types following the rule described in **Supporting Information (Section 3)** into 'D+' (positively charged), 'HBD', 'HYD' (hydrophobic), 'AR' (aromatic), 'A-' (negatively charged), 'HBA', 'MCD' (metal), with unclassifiable cells being assigned 'DEL'.

(3) The constructed grid was then transformed into an 11-channel 3D image, with each channel encoded a specific feature and cells marked as 0 or 1 to denote the absence or presence of the corresponding feature, respectively. A convolution kernel, as depicted in **Figure S12 C**, was used to process (stride 1, padding 3) all channels to produce a relatively continuous valued image. This image was used as the input of the target when training the binding affinity prediction model.

The preprocessing process and a case of generated grid were shown in **Figure S12 AB**.

A base framework of the binding affinity prediction model was shown in **Figure 8**. Specifically, the SE3CNN, a 3D steerable CNN, was employed to enhance the learning of rotationally equivariant features, as developed in Weiler's work[40]. Components delineated by orange boxes were designated for preprocessing and were not trained with the model. Attention block incorporated a straightforward application of cross-attention to amalgamate the target's and ligand's information. FC stood for fully connected networks. Ablation studies were conducted on the framework at four specific places, as indicated by numerical annotations in **Figure 8**. At mark '1', the SE3CNN was replaced with other complex CNN architectures as demonstrated in **Figure S13**, with the ablation referred to as 'moddedCNN'. At mark '2', the attention block was replaced with other three distinct fusion strategies: the direct addition of P_Emb and L_Emb following a layer normalization (deAttention0), the elimination of MatMul and SoftMax layers within the attention block and the direct addition of K, Q, V to fuse information to keep the parameter number levels (deAttention1), and the removal of redundant K-related layers in deAttention1 (deAttention2). At mark '3', an additional FC layer (512 to 512) following a LeakyReLU activation function was integrated to facilitate the learning of ligand representations during model training. At mark '4', the output was scaled and shifted to guide the model at the start of the training, by scaling 1.25 times then minus 2 (adjusted1), or by minus 2 (adjusted2). Dropout was also evaluated by adding three dropout layers subsequent to the SE3CNN, the first FC layer following L_Emb in the attention module, and the attention block, with the dropout rate set at 0.1/0.3/0.5. The three dropout layers were each in charge of processing information about the target, the ligand, and the fused information. Throughout the training and evaluation of models, the Adam optimizer was utilized, with a learning rate of 0.00003, weight decay of 0.01, batch size of 128, a random seed of 1, employing the MSELoss function, and conducting training 100 epochs (selecting the epoch yielding the best model performance for the final model). Following the extraction of 200 samples for the test set, the remaining dataset was used for 5-fold cross-validation training.

Baseline models served to demonstrate the efficacy of the final binding affinity prediction model. A 5-fold cross-validation approach was employed to identify the optimal baseline models. The Baseline SVR employed a support vector machine algorithm, exploring four kernel functions: 'linear', 'poly', 'rbf', and 'sigmoid'. The Baseline MLP was configured with the FC layer (11*14*14*14+512, 64), LeakyReLU, and FC layer (64, 1). Both baseline models utilized a concatenated representation of targets and ligands. The Baseline MLP_noProtein and Baseline MLP_noLigand were adjusted by excluding either the targets' or ligands' input, respectively, and modifying the dimensions of the input layer accordingly.

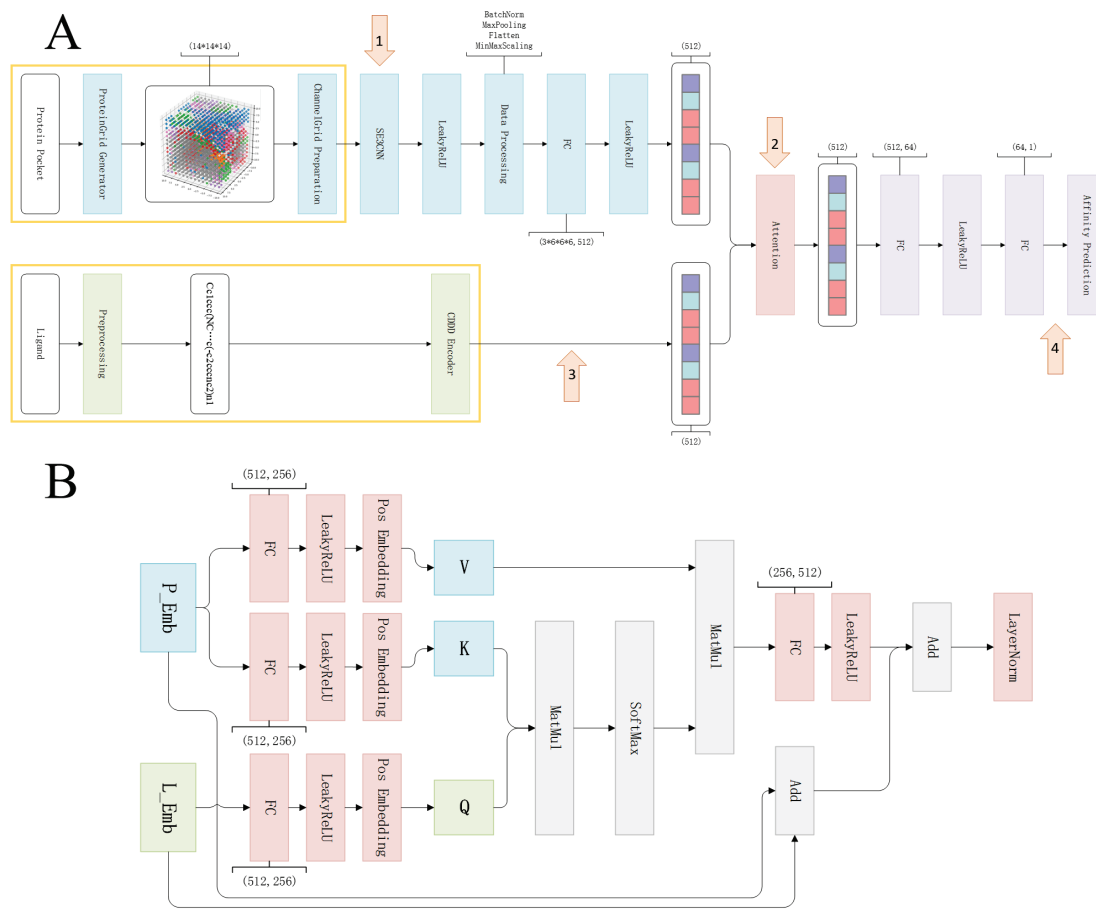

**Figure 8** The framework of the binding affinity prediction model. A: The model structure. B: The structure of the attention module.

The ADMET prediction model was developed and selected utilizing GridSearchCV methods in Scikit-learn. This process included five machine learning methods: MLP, GBDT, SVM, RF, KNN. The hyperparameters searched by GridSearchCV were detailed in **Table S6**.

**Molecule docking process**

In this study, we used GLIDE[41] in Schrodinger 2018 to perform all molecular dockings. The Protein Preparation Wizard was employed to preprocess the protein using default settings. Subsequently, the Virtual Screening Workflow was used to input SMILES of ligands and preprocess ligands using default settings, and generate grids centered on the original ligands (grid lengthis 13 Å). HTS (50% or all poses) and SP (100% best poses) were applied for the docking. For CMNPD virtual screening, HTS retained all poses for the top 100 molecules, whereas SP preserved the best poses for all molecules. Ligand Interaction was employed to predict the binding patterns.

# Acknowledgments

This work was supported by the "AI + Health Collaborative Innovation Cultivation" Project (Grant No. Z221100003522022), the National Key Research and Development Program (Grant No. 2022YFF1203003), the Peking University Medicine—StoneWise Joint Laboratory Project (Grant No. L202107), and the open fund of state key laboratory of Pharmaceutical Biotechnology, Nanjing University, China (Grant No. KF-202304). The authors declare that they have no competing interests.

# Reference

1  Wishart, D. S. *et al.* DrugBank 5.0: a major update to the DrugBank database for 2018. *Nucleic Acids Res* **46**, D1074-D1082 (2018). https://doi.org:10.1093/nar/gkx1037
2  Mendez, D. *et al.* ChEMBL: towards direct deposition of bioassay data. *Nucleic Acids Res* **47**, D930-D940 (2019). https://doi.org:10.1093/nar/gky1075
3  Ruddigkeit, L., van Deursen, R., Blum, L. C. & Reymond, J. L. Enumeration of 166 billion organic small molecules in the chemical universe database GDB-17. *J Chem Inf Model* **52**, 2864-2875 (2012). https://doi.org:10.1021/ci300415d
4  Mullard, A. The drug-maker's guide to the galaxy. *Nature* **549**, 445-447 (2017). https://doi.org:10.1038/549445a
5  Lipinski, C. F., Maltarollo, V. G., Oliveira, P. R., da Silva, A. B. F. & Honorio, K. M. Advances and Perspectives in Applying Deep Learning for Drug Design and Discovery. *Front Robot AI* **6**, 108 (2019). https://doi.org:10.3389/frobt.2019.00108
6  Zhang, Y. An In-depth Summary of Recent Artificial Intelligence Applications in Drug Design. arXiv:2110.05478 (2021). <https://ui.adsabs.harvard.edu/abs/2021arXiv211005478Z>.
7  Gupta, R. *et al.* Artificial intelligence to deep learning: machine intelligence approach for drug discovery. *Mol Divers* **25**, 1315-1360 (2021). https://doi.org:10.1007/s11030-021-10217-3
8  Sun, D., Gao, W., Hu, H. & Zhou, S. Why 90% of clinical drug development fails and how to improve it? *Acta Pharm Sin B* **12**, 3049-3062 (2022). https://doi.org:10.1016/j.apsb.2022.02.002
9  Brown, N. *et al.* Artificial intelligence in chemistry and drug design. *J Comput Aided Mol Des* **34**, 709-715 (2020). https://doi.org:10.1007/s10822-020-00317-x
10 Emmert-Streib, F., Yang, Z., Feng, H., Tripathi, S. & Dehmer, M. An Introductory Review of Deep Learning for Prediction Models With Big Data. *Front Artif Intell* **3**, 4 (2020). https://doi.org:10.3389/frai.2020.00004
11 Sejnowski, T. J. The unreasonable effectiveness of deep learning in artificial intelligence. *Proc Natl Acad Sci U S A* **117**, 30033-30038 (2020).

https://doi.org:10.1073/pnas.1907373117

12    Meyers, J., Fabian, B. & Brown, N. De novo molecular design and generative models. *Drug Discov Today* **26**, 2707-2715 (2021). https://doi.org:10.1016/j.drudis.2021.05.019

13    Devi, R. V., Sathya, S. S. & Coumar, M. S. Evolutionary algorithms for de novo drug design – A survey. *Applied Soft Computing* **27**, 543-552 (2015). https://doi.org:10.1016/j.asoc.2014.09.042

14    Martinelli, D. D. Generative machine learning for de novo drug discovery: A systematic review. *Comput Biol Med* **145**, 105403 (2022). https://doi.org:10.1016/j.compbiomed.2022.105403

15    Cheng, Y., Gong, Y., Liu, Y., Song, B. & Zou, Q. Molecular design in drug discovery: a comprehensive review of deep generative models. *Brief Bioinform* **22** (2021). https://doi.org:10.1093/bib/bbab344

16    Bilodeau, C., Jin, W. G., Jaakkola, T., Barzilay, R. & Jensen, K. F. Generative models for molecular discovery: Recent advances and challenges. *Wires Comput Mol Sci* **12** (2022). https://doi.org:ARTN e1608
10.1002/wcms.1608

17    Liu, N., Jin, H., Zhang, L. & Liu, Z. Plug-in Models: A Promising Direction for Molecular Generation. *Health Data Science* **3** (2023). https://doi.org:10.34133/hds.0092

18    Kotsias, P.-C. *et al.* Direct steering of de novo molecular generation with descriptor conditional recurrent neural networks. *Nature Machine Intelligence* **2**, 254-265 (2020). https://doi.org:10.1038/s42256-020-0174-5

19    Gupta, A. *et al.* Generative Recurrent Networks for De Novo Drug Design. *Mol Inform* **37** (2018). https://doi.org:10.1002/minf.201700111

20    Maziarka, L. *et al.* Mol-CycleGAN: a generative model for molecular optimization. *J Cheminform* **12**, 2 (2020). https://doi.org:10.1186/s13321-019-0404-1

21    Yoshizawa, T. *et al.* Selective Inhibitor Design for Kinase Homologs Using Multiobjective Monte Carlo Tree Search. *J Chem Inf Model* **62**, 5351-5360 (2022). https://doi.org:10.1021/acs.jcim.2c00787

22    Hoffman, S. C., Chenthamarakshan, V., Wadhawan, K., Chen, P.-Y. & Das, P. Optimizing molecules using efficient queries from property evaluations. *Nature Machine Intelligence* **4**, 21-31 (2021). https://doi.org:10.1038/s42256-021-00422-y

23    Mouchlis, V. D. *et al.* Advances in de Novo Drug Design: From Conventional to Machine Learning Methods. *Int J Mol Sci* **22** (2021). https://doi.org:10.3390/ijms22041676

24    Hartenfeller, M., Proschak, E., Schuller, A. & Schneider, G. Concept of combinatorial de novo design of drug-like molecules by particle swarm optimization. *Chem Biol Drug Des* **72**, 16-26 (2008). https://doi.org:10.1111/j.1747-0285.2008.00672.x

25    Winter, R. *et al.* Efficient multi-objective molecular optimization in a continuous latent space. *Chem Sci* **10**, 8016-8024 (2019). https://doi.org:10.1039/c9sc01928f

26    Winter, R., Montanari, F., Noe, F. & Clevert, D. A. Learning continuous and data-driven molecular descriptors by translating equivalent chemical representations. *Chem Sci* **10**, 1692-1701 (2019). https://doi.org:10.1039/c8sc04175j

27    James V. Miranda, L. PySwarms: a research toolkit for Particle Swarm Optimization in Python. *The Journal of Open Source Software* **3** (2018). https://doi.org:10.21105/joss.00433

28    Bottcher, J. *et al.* Fragment-based discovery of a chemical probe for the PWWP1 domain of NSD3. *Nat Chem Biol* **15**, 822-829 (2019). https://doi.org:10.1038/s41589-019-0310-x

29    Yun, C. H. *et al.* Structures of lung cancer-derived EGFR mutants and inhibitor complexes: mechanism of activation and insights into differential inhibitor sensitivity. *Cancer Cell* **11**, 217-227 (2007). https://doi.org:10.1016/j.ccr.2006.12.017

30    Brown, N., Fiscato, M., Segler, M. H. S. & Vaucher, A. C. GuacaMol: Benchmarking Models for de Novo Molecular Design. *J Chem Inf Model* **59**, 1096-1108 (2019). https://doi.org:10.1021/acs.jcim.8b00839

31    Liu, N. *et al.* Chemical Space, Scaffolds, and Halogenated Compounds of CMNPD: A Comprehensive Chemoinformatic Analysis. *J Chem Inf Model* **61**, 3323-3336 (2021). https://doi.org:10.1021/acs.jcim.1c00162

32    Lyu, C. *et al.* CMNPD: a comprehensive marine natural products database towards facilitating drug discovery from the ocean. *Nucleic Acids Res* **49**, D509-D515 (2021). https://doi.org:10.1093/nar/gkaa763

33    Couture, J. F., Hauk, G., Thompson, M. J., Blackburn, G. M. & Trievel, R. C. Catalytic roles for carbon-oxygen hydrogen bonding in SET domain lysine methyltransferases. *J Biol Chem* **281**, 19280-19287 (2006). https://doi.org:10.1074/jbc.M602257200

34    Hevener, K. E. *et al.* Structural studies of pterin-based inhibitors of dihydropteroate synthase. *J Med Chem* **53**, 166-177 (2010). https://doi.org:10.1021/jm900861d

35    Zhao, Y. *et al.* Structure-based design of novel pyrimido[4,5-c]pyridazine derivatives as dihydropteroate synthase inhibitors with increased affinity. *ChemMedChem* **7**, 861-870 (2012). https://doi.org:10.1002/cmdc.201200049

36    Liu, Z. *et al.* PDB-wide collection of binding data: current status of the PDBbind database. *Bioinformatics* **31**, 405-412 (2015). https://doi.org:10.1093/bioinformatics/btu626

37    Huang, K. *et al.* Artificial intelligence foundation for therapeutic science. *Nat Chem Biol* **18**, 1033-1036 (2022). https://doi.org:10.1038/s41589-022-01131-2

38    Ertl, P. & Schuffenhauer, A. Estimation of synthetic accessibility score of drug-like molecules based on molecular complexity and fragment contributions. *J Cheminform* **1**, 8 (2009). https://doi.org:10.1186/1758-2946-1-8

39    Desaphy, J., Azdimousa, K., Kellenberger, E. & Rognan, D. Comparison and druggability prediction of protein-ligand binding sites from pharmacophore-annotated cavity shapes. *J Chem Inf Model* **52**, 2287-2299 (2012). https://doi.org:10.1021/ci300184x

40    Weiler, M., Geiger, M., Welling, M., Boomsma, W. & Cohen, T. 3D steerable CNNs: Learning rotationally equivariant features in volumetric data. *Adv Neur In* **31** (2018).

41    Friesner, R. A. *et al.* Extra precision glide: docking and scoring incorporating a model of hydrophobic enclosure for protein-ligand complexes. *J Med Chem* **49**, 6177-6196 (2006). https://doi.org:10.1021/jm051256o

# Supporting Information

## 1 Supplementary figures

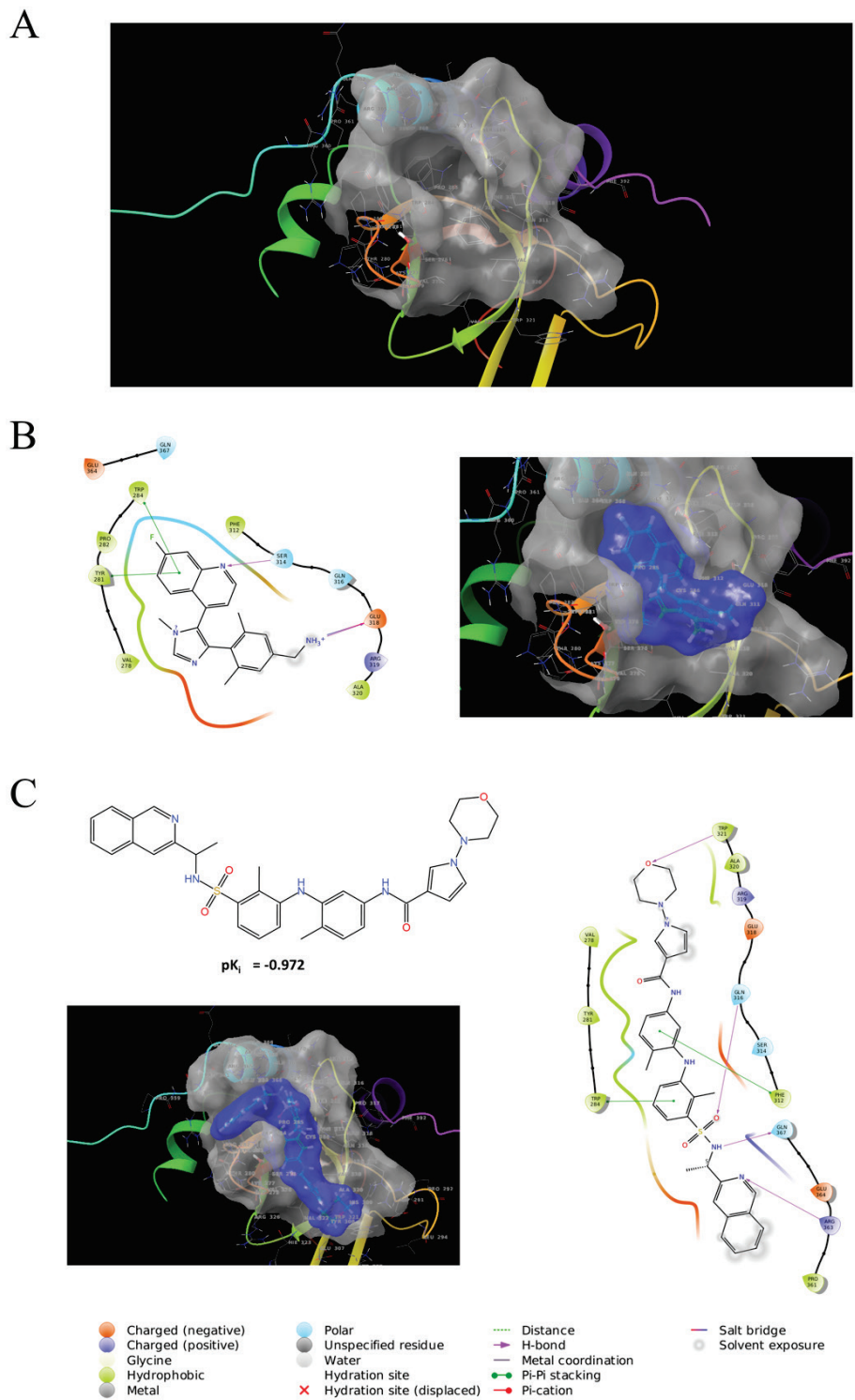

**Figure S1 Case study of single-objective molecule generation. A: Surface of the target binding site. B: Predicted binding mode and 3D spatial occupancy diagram for the original**

ligand of 6G2O. C: Structure of the generated case molecule. In the spatial occupancy diagrams, blue represents the ligand binding surface, and grey represents the target binding surface.

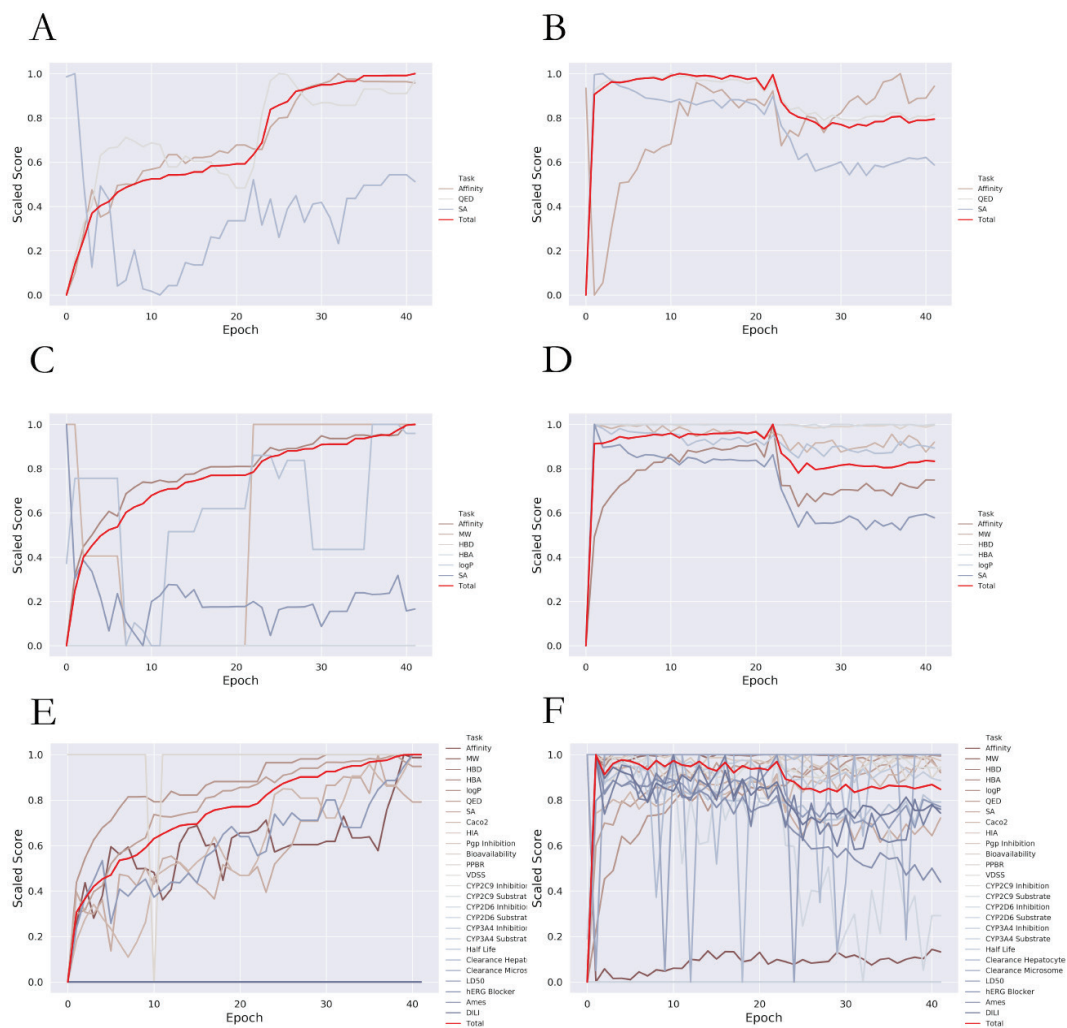

Figure S2 The multi-objective task scores during the iteration (binding to EGFR). A/C/E: gBest values (repeated experiments' average) for 3/6/26-objective tasks. B/D/F: Particles' average values (repeated experiments' average) for 3/6/26-objective tasks.

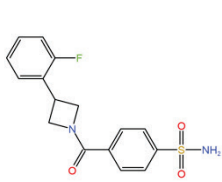
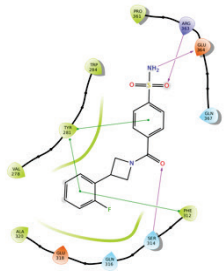
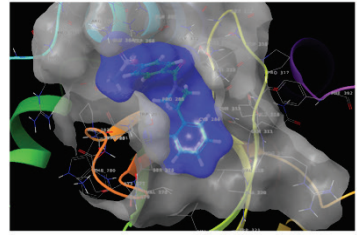

A

pKi = -3.434
QED = 0.928
SA = 1.961

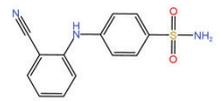
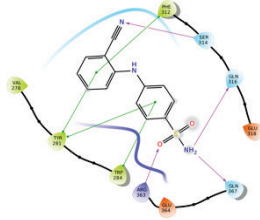
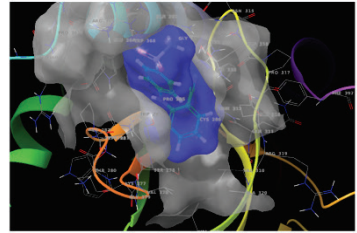

pKi = -3.899
QED = 0.893
SA = 1.867

B

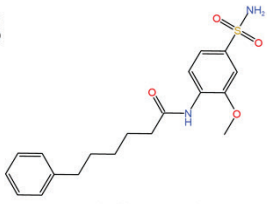
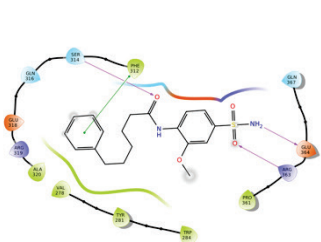
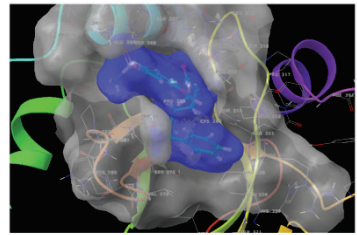

pKi = -3.403   HBA = 6
MW = 376.5     logP = 3.084
HBD = 3        SA = 1.849

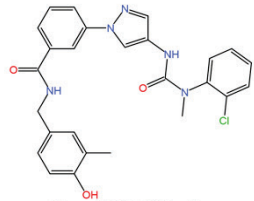
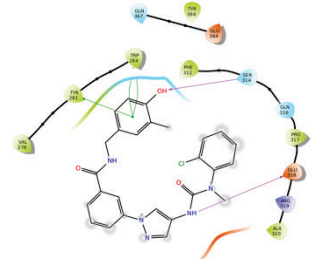
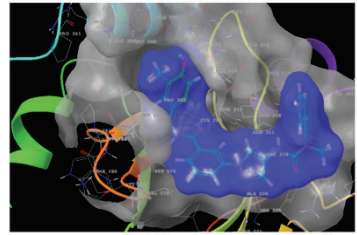

pKi = -1.586   HBA = 8
MW = 490.0     logP = 5.138
HBD = 3        SA = 2.491

C

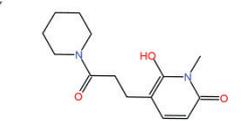
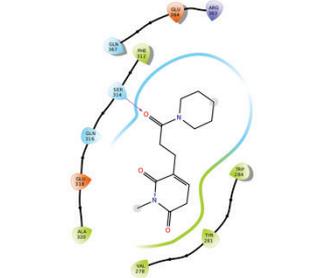
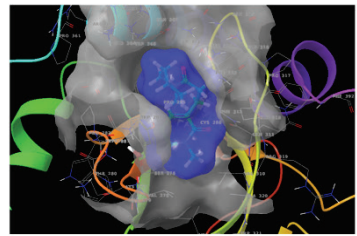

pKi = -5.857   logP = 1.036
MW = 264.3     QED = 0.887
HBD = 1        SA = 2.366
HBA = 5        ADMET SR = 100.0%

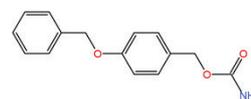
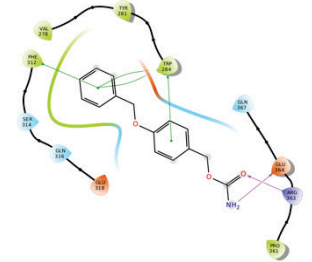
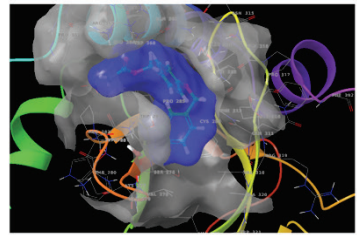

pKi = -6.244   logP = 2.861
MW = 257.3     QED = 0.895
HBD = 2        SA = 1.556
HBA = 4        ADMET SR = 100.0%

| | | | |
|---|---|---|---|
| ● Charged (negative) | ● Polar | ┄ Distance | — Salt bridge |
| ● Charged (positive) | ● Unspecified residue | ➤ H-bond | ◎ Solvent exposure |
| ● Glycine | ● Water | — Metal coordination | |
| ● Hydrophobic | ● Hydration site | ●— Pi-Pi stacking | |
| ● Metal | ✕ Hydration site (displaced) | ●— Pi-cation | |

**Figure S3** Case study of multi-objective molecule generation. A/B/C: Structures of the generated case molecules, predicted property values, predicted binding mode and 3D spatial occupancy diagram for 3/6/26-objective tasks.

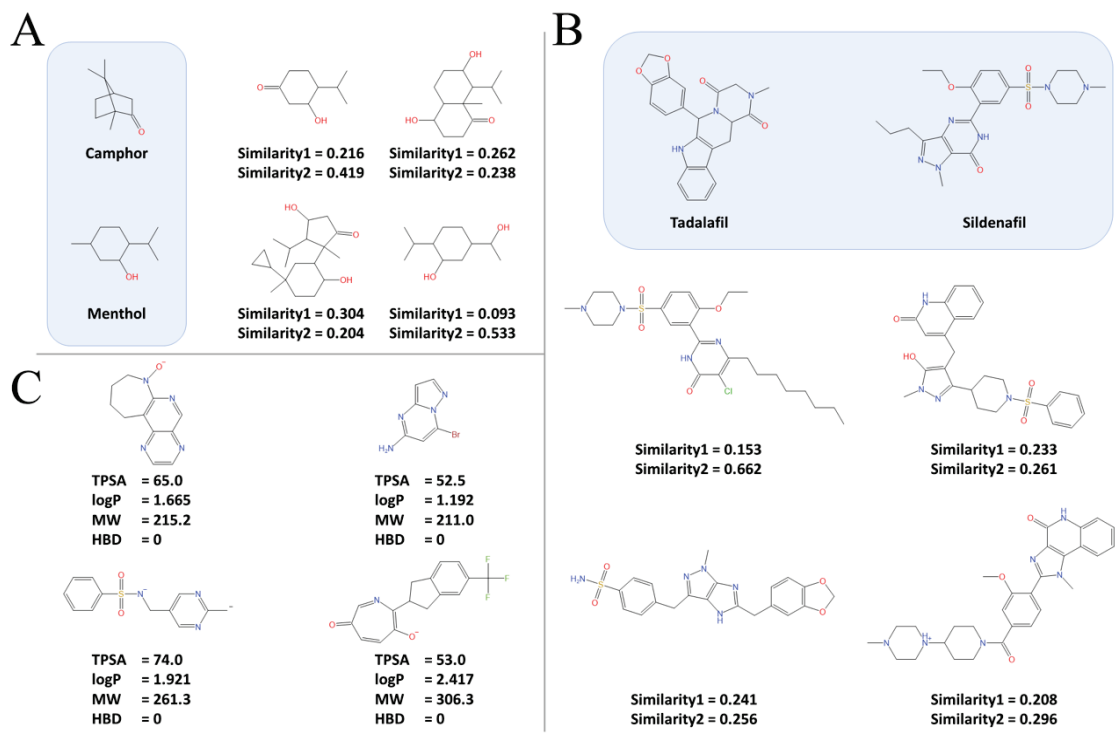

**Figure S4** Case molecules of baseline multi-objective tasks. A: Median 1; B: Median 2; C: CNS MPO. The molecules outlined in light blue represent the reference molecules in the Median task, while the rest are generated case molecules.

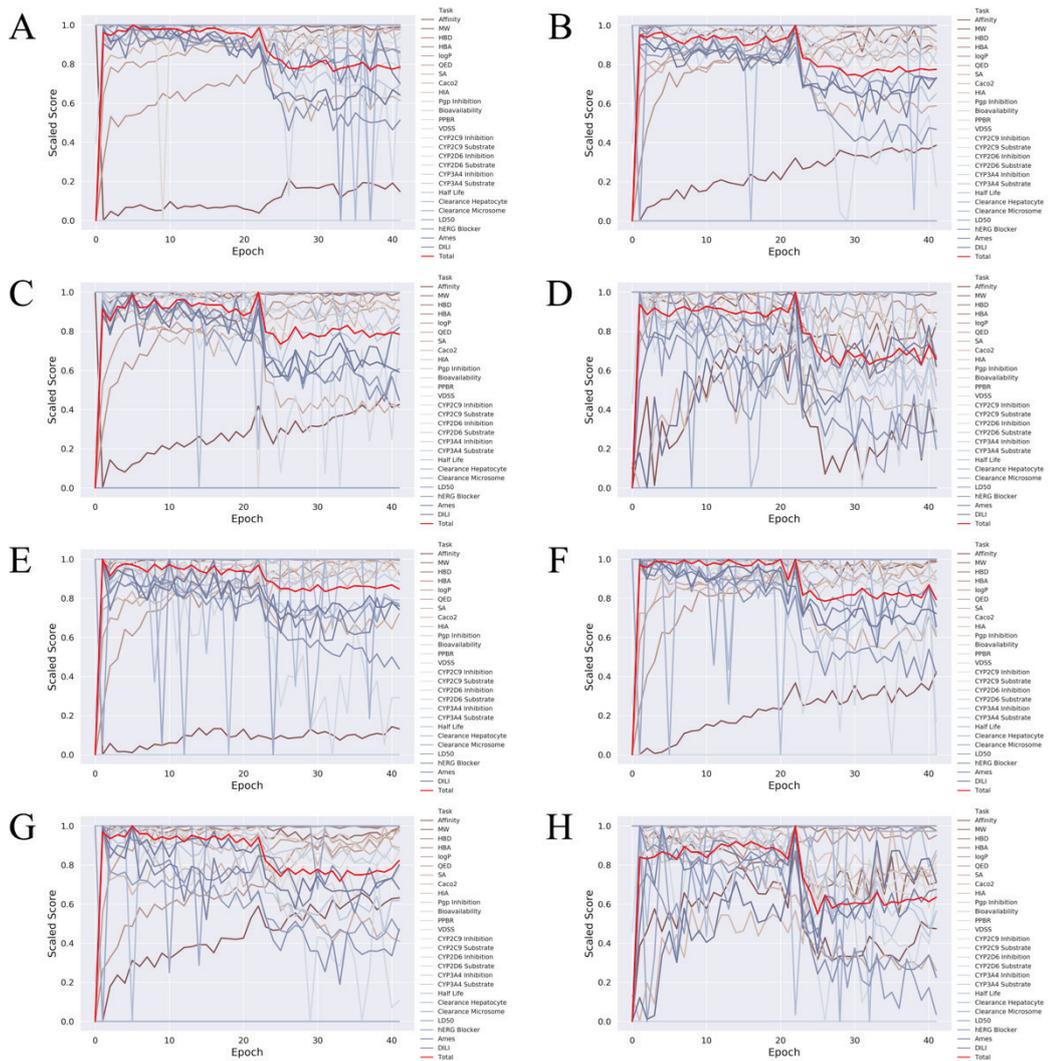

Figure S5 The 26-objective task scores during the iteration under different weight settings. A/B/C/D: Particles' average values (repeated experiments' average) when the binding affinity weights are set to 1.0/2.0/4.0/8.0 with NSD3 as the target. E/F/G/H: Similar curves for EGFR, with the same meanings and weight settings as mentioned above.

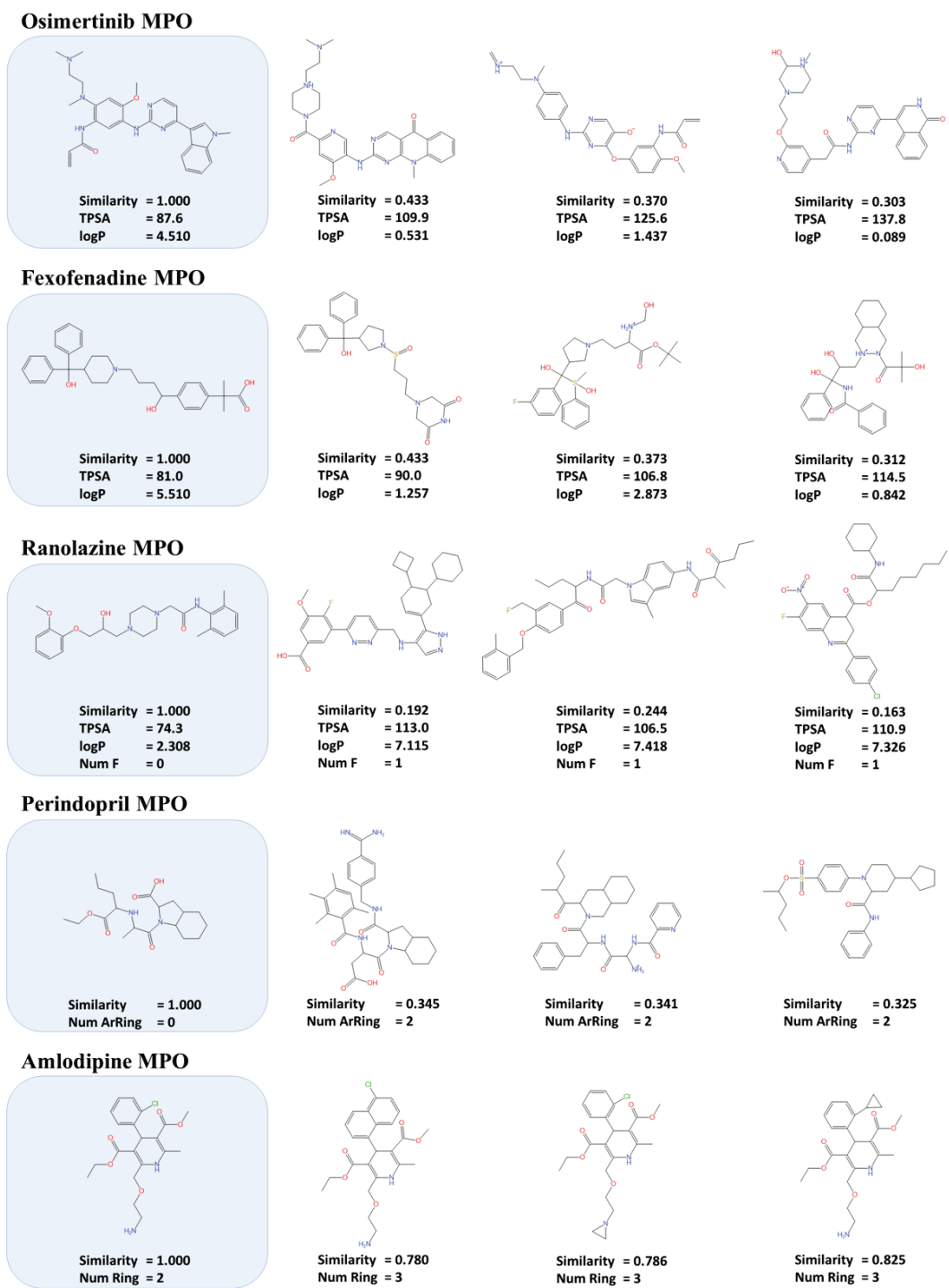

**Figure S6** The structures and predicted/computed properties of baseline structure optimization task case molecules. The molecules outlined in light blue represent reference molecules for each task, while the rest are generated case molecules.

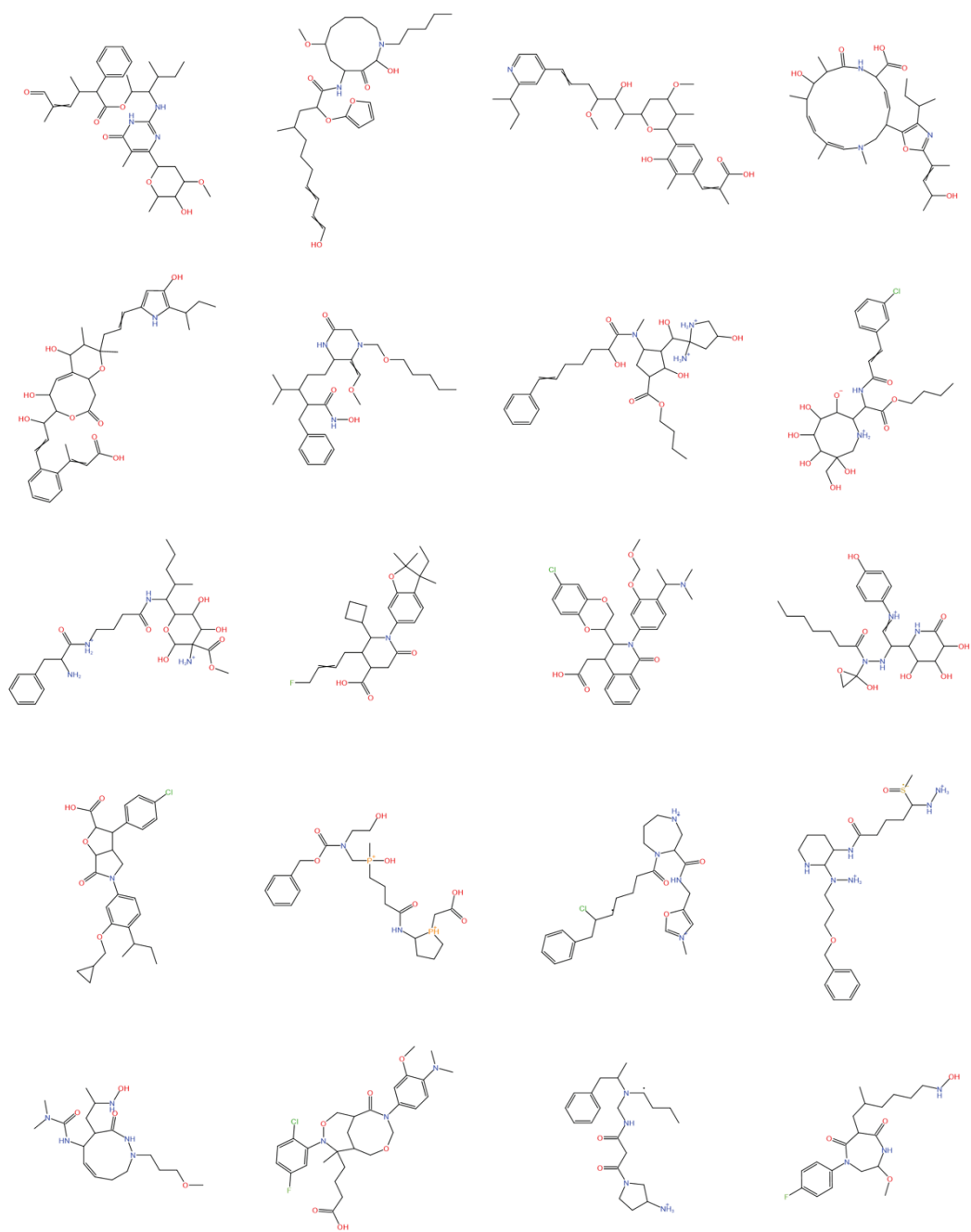

**Figure S7 Drug-like big MNPs analogues generation.**

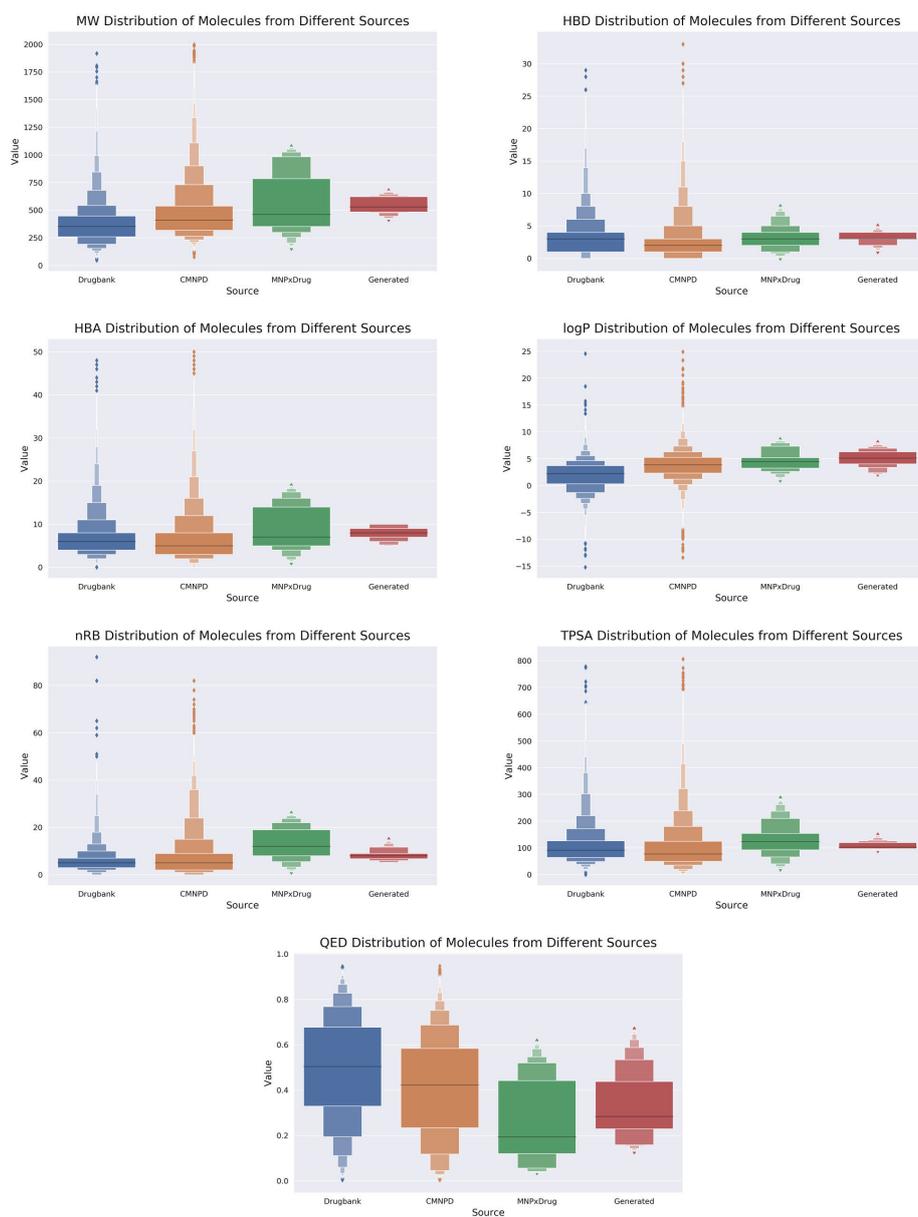

**Figure S8 Seven key drug-like properties of DrugBank, CMNPD, big MNPs in DrugBank (MNPxDrug), and generated molecules.**

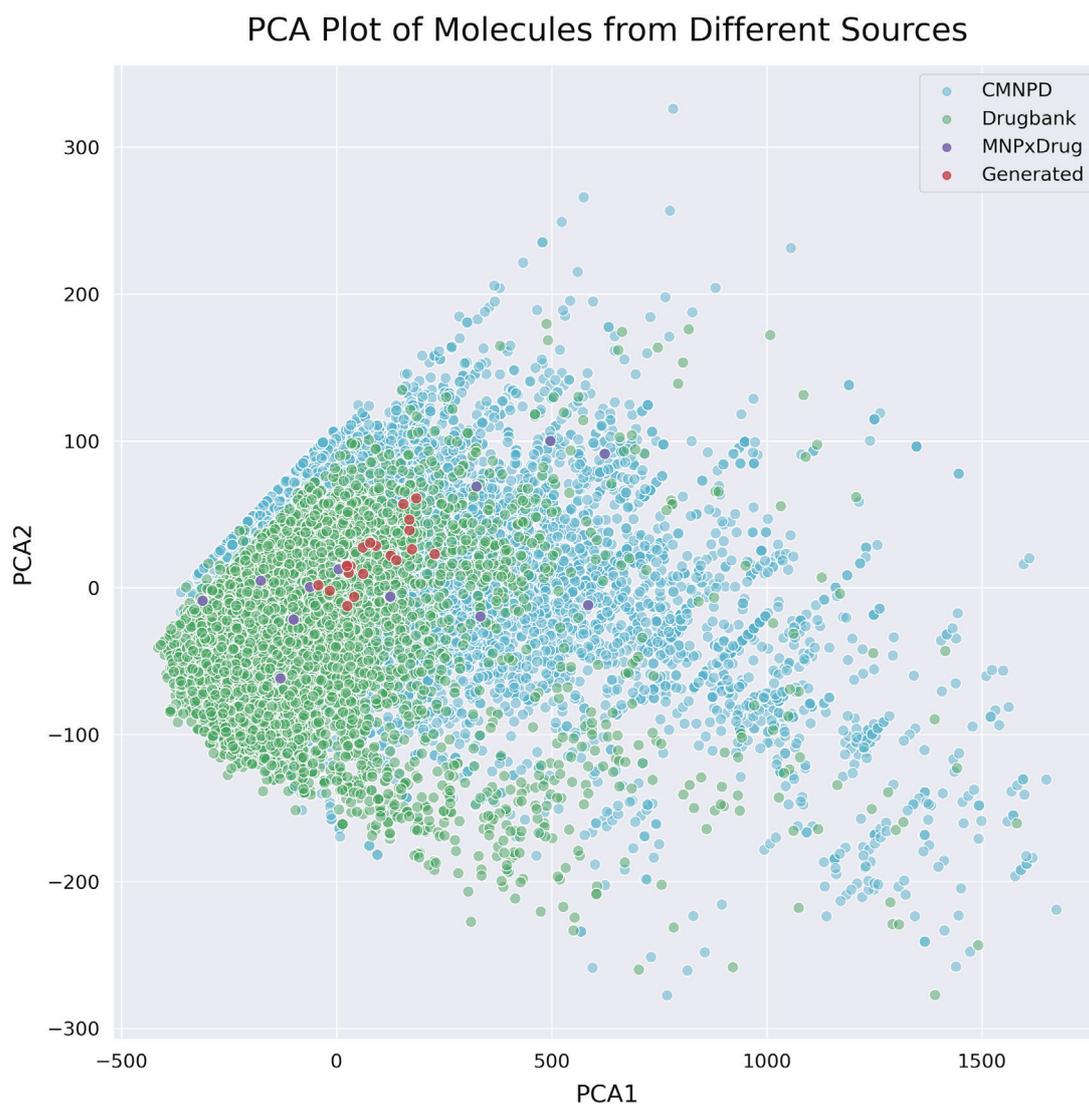

**Figure S9 PCA diagram of chemical space of DrugBank, CMNPD, big MNPs in DrugBank (MNPxDrug), and generated molecules.**

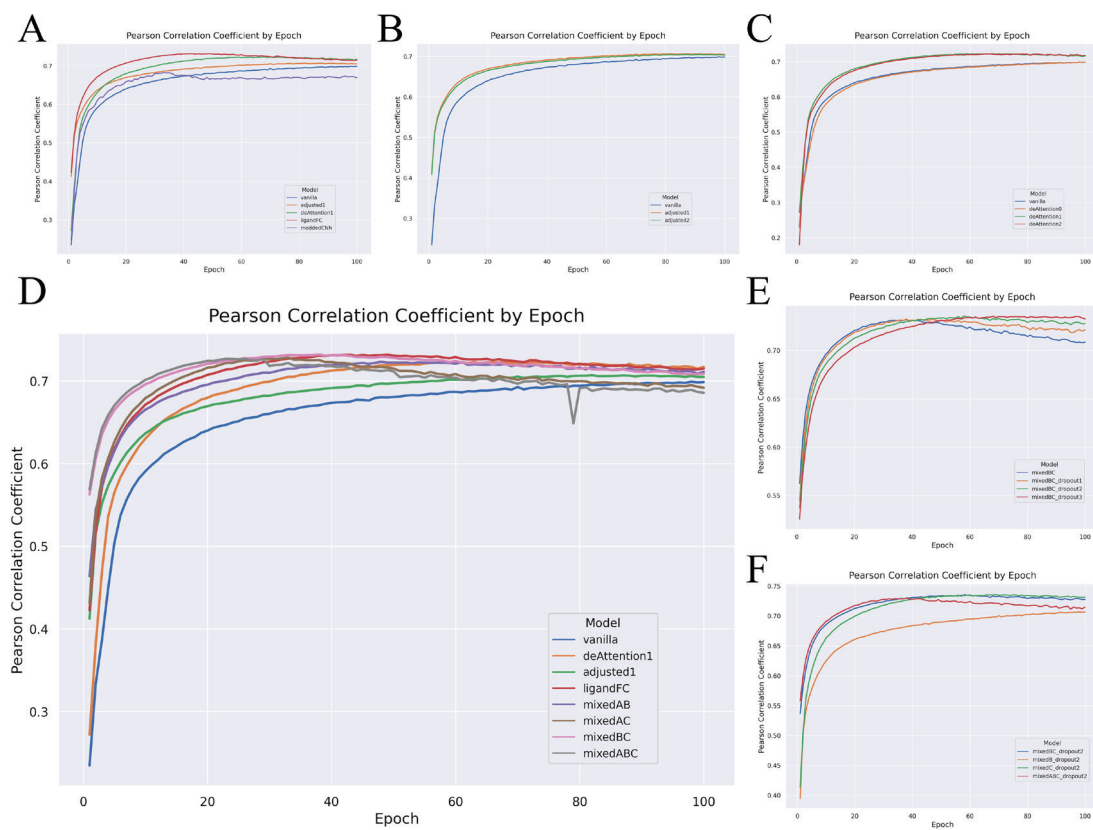

**Figure S10** The average pearson correlation coefficient of the target-based binding affinity prediction model varies with iteration. A: Comparison of the first round of model modifications; B: Comparison of different scaling offset methods; C: Comparison of different methods for removing attention modules; D: Comparative testing of the combination of modified modules; E: Dropout testing; F: The ablation experiment of the final model.

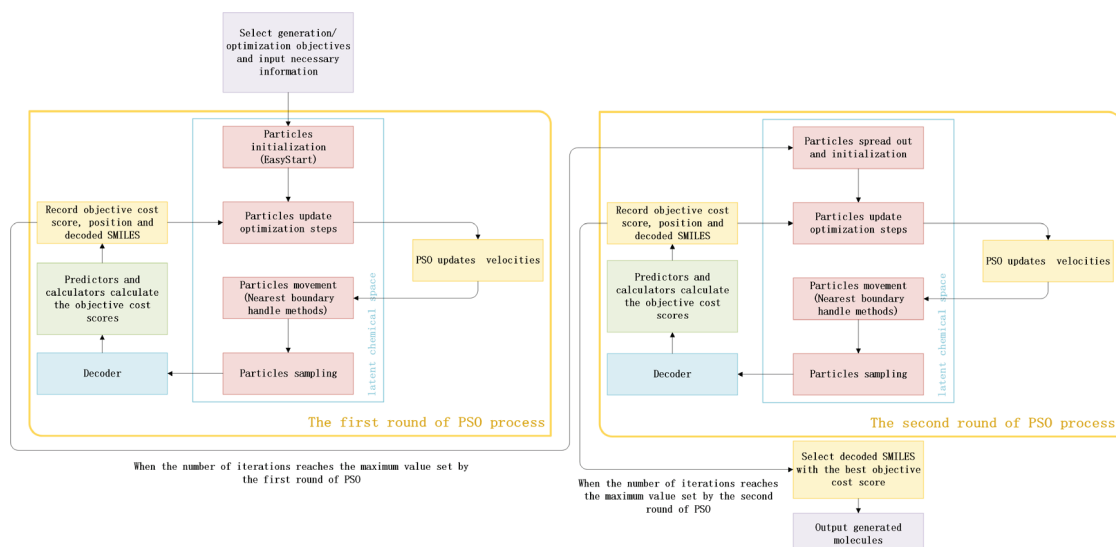

**Figure S11 The structures of the final Multi-functional plug-in molecule generation model PSO-ENP.**

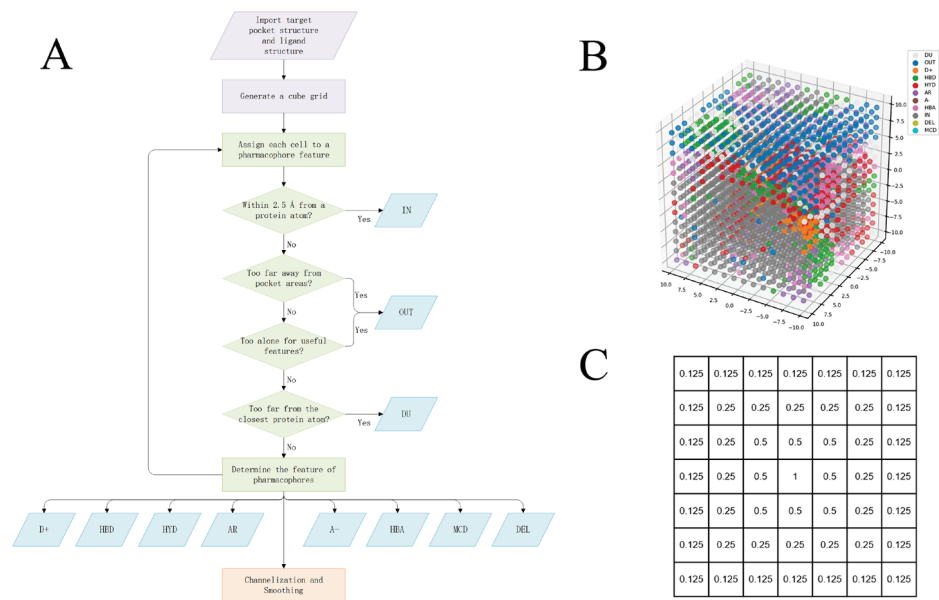

**Figure S12 Construction of cube grid with pharmacophore feature of target pocket. A: The preprocessing process; B: A case of generated grid before channelization and smoothing; C: 2D display of 3D convolutional core.**

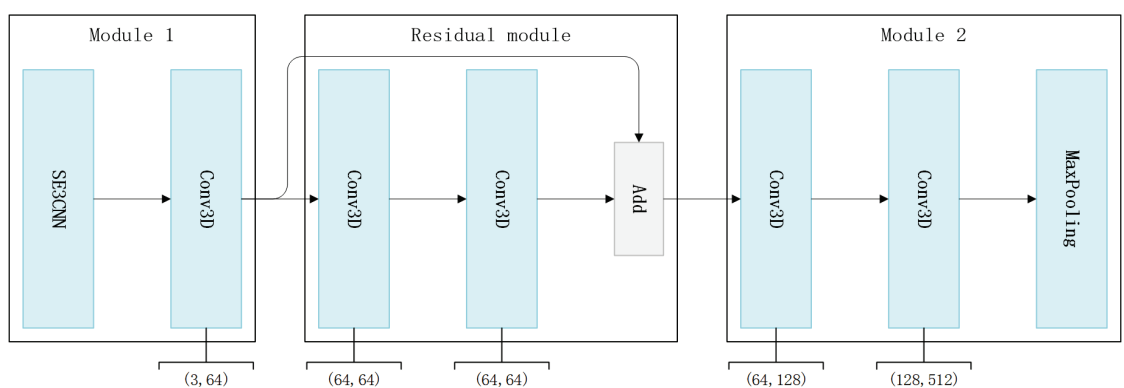

**Figure S13 CNN structures used to replace SE3CNN.**

# 2 Supplementary tables

Table S1 Baseline test results of single-objective molecule generation tasks. Bold numbers are best values of tasks.

| Task | Description | Graph MCT | PSO-ENP |
|---|---|---|---|
| Celecoxib Rediscovery | Rediscover Celecoxib | 0.355 | **1.000** |
| Troglitazone Rediscovery | Rediscover Troglitazone | 0.311 | **1.000** |
| Thiothixene Rediscovery | Rediscover Thiothixene | 0.311 | **0.461** |
| Aripiprazole Similarity | Rediscover Aripiprazole | 0.380 | **0.497** |
| Albuterol Similarity | Generate molecules similar to Albuterol | **0.749** | 0.688 |
| Mestranol Similarity | Generate molecules similar to Mestranol | 0.402 | **0.578** |
| logP 1 | Generate molecules with a logP of -1 | 0.986 | **0.997** |
| logP 2 | Generate molecules with a logP of 8 | 0.980 | **0.998** |
| TPSA | Generate molecules with a TPSA of 150 | **1.000** | **1.000** |
| QED | Generate molecules with high QED | 0.945 | **0.947** |

Table S2 Baseline test results of multi-objective molecule generation tasks. Bold numbers are best values of tasks.

| Task | Description | Graph MCT | PSO-ENP |
|---|---|---|---|
| Median 1 | Generate molecules with high similarity to both Camphor and Menthol | 0.225 | **0.250** |
| Median 2 | Generate molecules with high similarity to both Tadalafil and Sildenafil | 0.170 | **0.218** |
| CNS MPO | Generate molecules with a TPSA between 40 and 90, a logP less than 5, a molecular weight less than 360, and no HBDs | **1.000** | **1.000** |

**Table S3 Optimization rates and successful rates of 26-objective tasks.**

| Weight | Target | OR$_{best}$ Average% | OR$_{best}$ Half Rate% | OR$_{pop}$ Average% | OR$_{pop}$ Half Rate% | SR Average% |
|---|---|---|---|---|---|---|
| **1.0** | NSD3 | 76.2 | 90.0 | 91.7 | 100.0 | 94.8 |
|  | EGFR | 80.0 | 100.0 | 93.6 | 100.0 | 96.2 |
| **2.0** | NSD3 | 76.3 | 90.0 | 91.4 | 100.0 | 95.2 |
|  | EGFR | 71.4 | 100.0 | 92.5 | 100.0 | 96.2 |
| **4.0** | NSD3 | 65.7 | 80.0 | 90.6 | 100.0 | 94.8 |
|  | EGFR | 67.7 | 90.0 | 89.6 | 100.0 | 96.9 |
| **8.0** | NSD3 | 52.8 | 60.0 | 88.9 | 100.0 | 92.7 |
|  | EGFR | 59.5 | 75.0 | 88.7 | 100.0 | 95.4 |

**Table S4 Final performances of all 13 ADMET classification prediction models.**

| TDC name | Best Algorithm | Best combination of hyperparameters | Validation ACC | Validation AUC | Test ACC | Test AUC |
|---|---|---|---|---|---|---|
| hia_hou | SVM | {C: 10.0, kernel: poly} | 0.9067 | 0.9187 | 0.8291 | 0.9547 |
| pgp_broccatelli | SVM | {C: 1.0, kernel: rbf} | 0.8530 | 0.9292 | 0.8449 | 0.9180 |
| bioavailability_ma | SVM | {C: 10.0, kernel: poly} | 0.7832 | 0.7154 | 0.7422 | 0.7073 |
| bbb_martins | SVM | {C: 2.0, kernel: poly} | 0.8461 | 0.8779 | 0.8695 | 0.8805 |
| cyp2d6_veith | SVM | {C: 1.0, kernel: rbf} | 0.8553 | 0.8483 | 0.8774 | 0.8554 |
| cyp3a4_veith | SVM | {C: 2.0, kernel: rbf} | 0.8056 | 0.8862 | 0.7908 | 0.8830 |
| cyp2c9_veith | SVM | {C: 2.0, kernel: poly} | 0.7989 | 0.8687 | 0.8069 | 0.8822 |
| cyp2d6_substrate_carbonmangels | SVM | {C: 0.01, kernel: linear} | 0.7181 | 0.7101 | 0.7259 | 0.7600 |
| cyp3a4_substrate_carbonmangels | SVM | {C: 1.0, kernel: rbf} | 0.6542 | 0.7106 | 0.6296 | 0.6360 |
| cyp2c9_substrate_carbonmangels | KNN | {n_neighbors': 49, weights: distance} | 0.8071 | 0.6811 | 0.7185 | 0.5931 |
| herg | SVM | {C: 10.0, kernel: rbf} | 0.7820 | 0.8174 | 0.8409 | 0.7398 |
| ames | SVM | {C: 1.0, kernel: rbf} | 0.6568 | 0.7253 | 0.7632 | 0.8292 |
| dili | SVM | {C: 1.0, kernel: rbf} | 0.6992 | 0.7893 | 0.7500 | 0.8683 |

**Table S5 Final performances of all 7 ADMET regression prediction models.**

| TDC name | Best Algorithm | Best combination of hyperparameters | Validation Pearson | Validation R2 | Test Pearson | Test R2 |
|---|---|---|---|---|---|---|
| caco2_wang | SVM | {C: 2.0, kernel: rbf} | 0.7281 | 0.5102 | 0.6426 | 0.3506 |
| ppbr_az | SVM | {C: 10.0, kernel: rbf} | 0.5409 | 0.2061 | 0.5165 | 0.2106 |
| vdss_lombardo | SVM | {C: 10.0, kernel: poly} | 0.3757 | 0.1366 | 0.3308 | 0.1065 |
| half_life_obach | KNN | {n_neighbors: 19, weights: distance} | 0.2290 | 0.0992 | 0.2689 | 0.0649 |
| clearance_microsome_az | RF | {max_features: 32, min_samples_split: 4} | 0.4367 | 0.1895 | 0.4480 | 0.1960 |
| clearance_hepatocyte_az | RF | {max_features: 128, min_samples_split: 3} | 0.4223 | 0.1723 | 0.2623 | 0.0193 |
| ld50_zhu | SVM | {C: 2.0, kernel: rbf} | 0.6384 | 0.4071 | 0.5745 | 0.2892 |

**Table S6 The hyperparameters of five machine learning methods included: MLP, GBDT, SVM, RF, KNN.**

| Methods | Hyperparameter | Values |
|---|---|---|
| MLP | activation | [relu, logistic, tanh, softmax] |
| | hidden_layer_sizes | [(256), (128), (64), (32), (16), (256, 64), (128, 32), (64, 16)] |
| | random_state | 1 |
| GBDT | max_depth | [3, 5, 7, 9] |
| | min_samples_split | [1, 2, 3, 4] |
| | max_features | [0.5, 0.8, 1.0] |
| | n_estimators | 200 |
| | random_state | 1 |
| SVM | C | [0.01, 0.1, 1.0, 2.0, 10.0] |
| | kernel | [linear, rbf, sigmoid, poly] |
| | random_state | 1 |
| RF | min_samples_split | [1, 2, 3, 4] |
| | max_features | [round(0.5*n_features$^{0.5}$), round(1*n_features$^{0.5}$), round(2*n_features$^{0.5}$), round(3*n_features$^{0.5}$), round(4*n_features$^{0.5}$), round(5*n_features$^{0.5}$)] |
| | n_estimators | 20 |
| | random_state | 1 |
| KNN | n_neighbors | [1, 2, ..., 50] |
| | weights | [uniform, distance] |

For the RF models **max_features** parameter, n_features refers to the number of features in the input data, which is 1024 in this article; The round function refers to rounding to the nearest whole number.

**3 Pharmacophore feature allocation rules.**

**The following is the rule code for assigning pharmacophore features in Python, with variable names explained at the end of the code:**

```
if check_symbol==H:
    if check_name!=H:
        if check_resn in [LYS]:
            return D+
        elif check_resn == ARG:
            if check_name!=HE:
                return D+
            else:
                return HBD
        else:
            return HBD
    else:
        return HBD
elif check_resn in metalGear:
    return MCD
else:
    try:
        return aaBigDict[check_resn][check_name]
    except:
        return DEL
```

**Check_symbol, check_name, and check_resn are the symbol (atomic label), name (atomic name), and resn (residue) of protein atoms obtained by reading the PDB file through the**

**Python library pymol2.**

**The settings for metalGear are as follows:**

metalGear =[

ZN,MG,CA,NA,NI,FE2,CU,K,FE,MN,CO,HG,CS,CD,SR,CAF,CAS,RB,CU1,

GA

]

**The settings for aaBigDict are as follows:**

Table S7 Settings for aaBigDict. Each row represents a different amino acid residue, listing the property classifications for each atom type within that residue.

| Residue | HYD | HBD | HBA | AR | A- | D+ |
|---|---|---|---|---|---|---|
| GLY | C, CA | N | O | | | |
| ALA | C, CA, CB | N | O | | | |
| VAL | C, CA, CB, CG1, CG2 | N | O | | | |
| LEU | C, CA, CB, CG, CD1, CD2 | N | O | | | |
| ILE | C, CA, CB, CG1, CG2, CD1 | N | O | | | |
| PHE | C, CA, CB | N | O | CG, CD1, CD2, CE1, CE2, CZ | | |
| PRO | C, CA, CB, CG, CD | | N, O | | | |
| SER | C, CA, CB | N, OG | O | | | |
| THR | C, CA, CB, CG2 | N, OG1 | O | | | |
| HIS | C, CA, CB, CG, CD2, CE1 | N, ND1 | O, NE2 | | | |
| TRP | C, CA, CB, CG, CD1, CD2 | N, NE1 | O | CE2, CE3, CZ2, CZ3, CH2 | | |
| CYS | C, CA, CB | N, SG | O | | | |
| ASP | C, CA, CB, CG | N | O, OD1 | | OD2 | |
| GLU | C, CA, CB, CG, CD | N | O, OE1 | | OE2 | |
| LYS | C, CA, CB, CG, CD, CE | N | O | | | NZ |
| TYR | C, CA, CB | N, OH | O | CG, CD1, CD2, CE1, CE2, CZ | | |
| MET | C, CA, CB, CG, CE | N | O, SD | | | |

| | | | | |
|---|---|---|---|---|
| ASN | C, CA, CB, CG | N, ND2 | O, OD1 | |
| GLN | C, CA, CB, CG, CD | N, NE2 | O, OE1 | |
| ARG | C, CA, CB, CG, CD, CZ | N, NE, NH2 | O | NH1 |